\documentstyle[multicol,tighten,aps,epsf]{revtex} 

\def\win{{W}} 
\def\px{{\bf x}} 
\def\py{{\bf y}} 
\def\pp{{\bf p}} 
\def\pq{{\bf q}}

\def\pP{{\bf P}} 
\def\pn{{\bf n}}
\def\pK{{\bf K}}
\def\pu{{\bf u}}

\begin{document} 
 
\title{Semiclassical Field Theory Approach to Quantum Chaos} 
\author{A. V. Andreev$^{a,b}$, B. D. Simons$^c$, O. Agam$^b$, and B. L.  
Altshuler$^b$} 
\address{$^a$Department of Physics, Massachusetts Institute of Technology, 77  
Massachusetts Avenue, Cambridge, MA\ 02139, USA\\ 
$^b$NEC Research Institute, 4 Independence Way, Princeton, NJ 08540, USA\\
$^c$Cavendish Laboratory, Madingley Road, Cambridge, CB3\ 0HE, UK} 
 
\maketitle 
 
\begin{abstract} 
We construct a field theory to describe energy averaged quantum statistical
properties of systems which are chaotic in their classical limit. An expression
for the generating function of general statistical correlators is presented
in the form of a functional supermatrix nonlinear $\sigma$-model where the
effective action involves the evolution operator of the classical dynamics.
Low-lying degrees of freedom of the field theory are shown to  
reflect the irreversible classical dynamics describing relaxation of
phase space distributions. The validity of this approach is investigated over
a wide range of energy scales. As well as recovering the universal long-time
behavior characteristic of random matrix ensembles, this approach accounts
correctly for the short-time limit yielding results which agree with the
diagonal approximation of periodic orbit theory. 
\end{abstract} 
 
\pacs{PACS numbers: 73.20.Dx, 73.20.Fz, 05.45.+b}  
 
\par 
\section{Introduction} 

The quantum description of systems which are chaotic in the classical limit 
is the subject of ``quantum chaos''. A wide variety of physical systems fall 
into this category. Amongst those most commonly studied are the neutron 
resonances of atomic nuclei~\cite{Porter65}, Rydberg atoms in strong magnetic 
fields~\cite{Delandelh}, and electrons in weakly disordered metallic grains 
(quantum dots)~\cite{LesHouches89}. The energy spectrum as a whole is 
specific for each individual chaotic system.  However, in contrast to 
integrable systems, each eigenstate is characterized only by its energy 
rather than by a set of quantum numbers~\cite{Gutzwiller90}. 
The variables in the corresponding Schr\"odinger equation do not 
separate, and an analytical solution is prohibited. Therefore, 
a useful description of highly excited eigenstates 
of chaotic systems is a statistical one. 

The statistical approach assumes certain averaging. Sometimes, as with
disorder, one can think about an ensemble of chaotic systems. 
In such cases ensemble averaging is sufficient. For an individual 
chaotic system, such as a Rydberg atom in a magnetic field, 
averaging over a wide energy interval is the only choice. 

Quantities investigated in the statistical approach to quantum 
spectra include various correlators of density of 
states (DoS) $\nu(E)={\rm Tr}\ \delta (E-\hat{H})$,  
where $\hat{H}$ is the Hamiltonian of the system.
Here, it is natural to measure energy differences 
in units of the mean level spacing $\Delta$.
Perhaps the property most frequently studied is the dimensionless 
two-point DoS correlator,
\begin{equation}
R_2(s)=\Delta^2 \left\langle \nu\left(E+{s\over 2}\Delta\right) 
\nu\left(E-{s\over 2}\Delta\right)\right\rangle-1, \label{R2}
\end{equation}
where $s$ is the dimensionless energy difference. As mentioned above, 
for disordered metals the statistical average, denoted by $\langle \cdots 
\rangle$ can be performed over different realizations of the random 
Hamiltonian while, in general, the average can be taken 
over a wide energy band.

Associated with each particular chaotic system there are 
typically two relevant energy scales. The first, $E_c$, is associated 
with the {\em classical} time scale $\tau_c=\hbar/E_c$ on which a  
density distribution in phase space becomes ergodic, i.e. spreads 
uniformly over the constant  energy shell. On time scales larger
than $\tau_c$, time averages over a classical trajectory can be substituted 
by microcanonical averages over the energy shell in phase space.  
In a cavity in which a quantum particle scatters ballistically from 
a boundary, the ``chaotic billiard'', the energy scale 
$E_c$ is typically set by the frequency of the shortest periodic 
orbit, or the inverse flight time across the system. In a 
weakly disordered metallic grain, on the other hand, the classical energy 
scale is set by the inverse transport time, or Thouless energy $E_c= \hbar D/
L^2$, where $D$ denotes the classical diffusion constant, and $L$ represents 
the system size. The second energy scale is set by the mean energy level 
spacing $\Delta$ which  defines the Heisenberg time $\tau_H=\hbar/\Delta$. 

The two energy scales can be combined into the dimensionless 
ratio, 
\begin{equation}
g=E_c/\Delta 
\end{equation}
which represents the ``dimensionless conductance''~\cite{Thouless}
of a chaotic system. The ergodic time $\tau_c=\hbar /E_c$ sets the 
scale beyond which the details of the classical dynamics of the 
system become irrelevant (as long as the system is chaotic). 
Correspondingly, for energy scales $s \ll g$, spectral 
statistics become universal, independent of the details of the 
underlying classical dynamics. To a very good approximation they are 
determined only by the global symmetries of the 
system~\cite{Bohigas84,Berry85,Les-Houches} such as 
T-invariance, and coincide with the universal Wigner-Dyson 
level statistics of the corresponding random matrix 
ensembles~\cite{Dyson}: Unitary (broken T-invariance), Orthogonal 
(spinless T-invariant systems), or Symplectic (T-invariant systems with 
spin-orbit interaction)~\cite{Mehta91,Haake}.
In the semiclassical limit, and in dimensions greater than one, 
the dimensionless conductance is large, $g\gg 1$, 
and universal statistics apply over a wide energy interval. 

The study of the universal and non-universal statistical properties of 
quantum chaotic systems has been conducted largely along two separate lines: 

\begin{enumerate}

\item
The first approach has been based on the nonlinear $\sigma$-model
proposed by Wegner~\cite{Wegner79}, and is applied to
the study of an {\em ensemble} of similar systems such as 
weakly disordered metallic grains in which electrons 
experience scattering by a random potential. Here, ensemble 
averaging is a crucial step exploited at an early stage of calculation.
The supersymmetric version of the nonlinear $\sigma$-model proposed 
by Efetov~\cite{Efetov82} provided a microscopic justification 
for the random matrix theory (RMT) description of universal long-time 
properties of such systems~\cite{Efetov83,Verbaarschot85}. The same theory
accounts for non-universal features of spectral statistics associated with 
the diffusive dynamics of electrons. However, such an approach suffers 
from two drawbacks: Firstly, it relies on the very existence 
of an ensemble. Very often we are concerned with non-stochastic 
chaotic systems, such as a chaotic billiard, 
where the notion of an ensemble is inappropriate. Secondly, this type 
of averaging tends to erase information about individual features 
of the system.

\item
The second approach has been based on Gutzwiller's trace formula
in which the semiclassical DoS is expressed as 
a sum over the classical periodic orbits \cite{Gutzwiller90}. 
This approach focuses on the behavior of {\em individual} systems, 
and statistical properties reply implicitly on 
energy averaging. This approach has proved to be a powerful 
tool in describing non-universal properties associated with short-time 
behaviour. However, its success in reproducing universal long 
time properties associated with RMT has been 
limited~\cite{Berry85,Wilkinson88}. 
In particular, it fails to account for the correct behavior at 
times in excess of the inverse level spacing or the Heisenberg time 
$\tau_H=\hbar/\Delta$. Thus, in spite of extensive numerical 
evidence~\cite{Casati80,Bohigas84} which supports the random matrix 
description of spectral statistics of {\em individual} systems at 
small energies, the origin of its success has remained obscure. 

\end{enumerate}

In Ref.~\cite{Andreev95} it was conjectured that the $\sigma$-model 
approach to diffusive systems could be generalised to include 
the wider class of chaotic systems.
In a recent and insightful development, Muzykantskii and 
Khmel'nitskii~\cite{Muzykantskii95} proposed a $\sigma$-model to 
describe short-time ballistic dynamics in 
disordered conductors. Although their argument relied solely on 
impurity averaging, they conjectured that their $\sigma$-model 
should apply even in the absence of disorder. 

Here we develop a semiclassical field theoretic description of
quantum spectral statistics of individual chaotic systems based 
{\em solely on energy averaging}. This method offers a  semiclassical 
description of  the statistical properties of individual quantum 
chaotic systems which accounts both for universal as well as 
non-universal features. The basic ingredients of the underlying 
classical dynamics are {\em no longer} the individual periodic
orbits, but general properties of the classical flow in phase space. 
It will be shown the the $\sigma$-model constructed in this way indeed
coincides with that proposed in Ref.~\cite{Muzykantskii95}.

Before introducing the main conclusions of this study, we begin by identifying
the questions which will be of most concern. To do so it is convenient to
draw on the insight offered by the study of the dynamics of a particle moving
in a background of weak randomly distributed scattering impurities. Amenable 
to the method of ensemble averaging, properties of this chaotic quantum 
mechanical problem are now reasonably well understood. 
 
What is the quantum evolution of a wavepacket in a background of impurities? 
According to the time of evolution, the dynamics of the wavepacket is
characterised by several quite distinct regimes. On time  
scales $t$ in excess of the mean free scattering time $\tau$,
the initial {\em ballistic} evolution of the wavepacket becomes
diffusive. At longer times the interference of different semiclassical  
paths induces a {\em quantum renormalization} of the bare diffusion constant  
$D=v_F^2\tau/d$. This leads to the phenomena of ``weak localisation'' and is 
responsible for the quantum coherence effects observed in transport  
properties of mesoscopic metallic conductors. If the impurity  
potential is not strong enough to localise the wavepacket altogether, the  
wavepacket continues to spread. After a time $\tau_c=L^2/D$, the typical  
transport or diffusion time, the wavepacket is spread approximately uniformly  
throughout the system. Further evolution of the wavepacket is therefore  
said to be {\em ergodic}. Beyond the ergodic time $t\gg \tau_c$ the evolution 
of the wavepacket becomes {\em universal}, independent of the individual  
features of the system. Finally, the spectral rigidity characteristic of  
quantum chaotic systems leads to an approximately coherent superposition  
or ``echo''~\cite{Prigodin94} of the wavepacket at $t=\tau_H$ after which
the wavepacket relaxes to a uniform distribution. 

A similar question can be asked about the quantum evolution of a wavepacket  
introduced into, say, an irregular cavity (quantum billiard) without  
impurities. In such systems it is widely believed that there too exists some  
ergodic time $\tau_c$ after which properties of the system become universal.  
However, at shorter time scales, how is the unstable nature of the classical  
dynamics reflected in the quantum evolution? Is there an analogue of  
quantum renormalization? What, in general, plays the role of  
the diffusion operator in describing the low energy degrees of freedom?  

In a recent study by three of us~\cite{Agam95}, a comparison of results taken 
from the leading order of diagrammatic perturbation theory for disordered 
metals within the diagonal approximation of periodic orbit theory 
led to the conjecture that, for general chaotic systems, the role of the 
diffusion operator is, in general, played by the generator $\hat{\cal L}$ 
of the classical evolution (or Perron-Frobenius) operator, 
$e^{-\hat{\cal L}t}$. If $\rho(\px,0)$ is an 
initial  probability density distribution defined as a function of 
phase space variables $\px\equiv(\pq,\pp)$, where $\pq$ and $\pp$ are
coordinates of the position and momentum vectors respectively, the density at 
a later time $t$ is given by 
\begin{equation} 
\rho(\px,t)=  e^{-\hat{\cal L}t} \rho(\px,0)\equiv  
\int_\Gamma d\py\ \delta[\px-\pu^t(\py)]\ \rho(\py,0), 
\label{pfoperator} 
\end{equation} 
where $\pu^t(\py)$ is the solution of classical equations of motion with 
initial conditions $\py$ at $t=0$, and $\Gamma$ covers the region of  
available phase space.  One eigenvalue of the Perron-Frobenius operator
is always unity and corresponds to the ergodic state. The rest of the 
eigenvalues are of the form $e^{-\gamma_{\mu}t}$, where 
$\Re (\gamma_{\mu})>0$, and appear as complex conjugate pairs. They are 
associated with decaying modes in which a smooth distribution relaxes into
the uniform ergodic state and are analogous to the diffusion 
modes of disordered systems. 
This analogy, therefore, suggests that the ergodic time in general chaotic 
systems is set by the first non-zero eigenvalue $\tau_c\sim \hbar/
\Re (\gamma_1)$. 

As well as confirming the conjecture made in Ref.~\cite{Agam95}, the field
theoretic approach will reveal that, in the semiclassical limit, all 
statistical spectral properties of the quantum system depend only on 
properties of the Perron-Frobenius operator. In particular, by considering 
only its leading eigenvalue $\gamma_0=0$, one exactly reproduces RMT, while 
taking into account higher modes enables one to characterize deviations from 
universality. The field theoretic approach provides for the first time a 
systematic and controlled way to investigate quantum corrections which lie 
beyond the diagonal approximation typically employed in the periodic orbit 
theory. 

The range of energy scales over which this approach is valid is illustrated 
in Fig.~1. Energy averaging will be performed over a wide energy band of 
width $W$ centered at $E_0$. As well as requiring that $E_0 \gg W$, we will 
assume that $W$ is much larger than the energy scale set by the first non-zero 
eigenvalue of $\hat{\cal L}$, $\gamma_1$. This is to ensure that the time 
scale $\hbar/W$ is fine enough to resolve the behavior of the classical 
dynamics over a time interval smaller than the ergodic time $\tau_c\sim 
\hbar/\gamma_1$. It will be also assumed that the finest energy scale, the 
mean level spacing $\Delta$, is much smaller than $\gamma_1$. Therefore, 
we will focus on a range of energy scales where $\Delta \ll \gamma_1 \ll W 
\ll E_0$.

As well as describing the universal regime of RMT, corresponding to energy
scales comparable to the mean level spacing $\Delta$, the field theory 
developed here properly describes non-universal behavior which appears at 
larger energy scales. However, to avoid encountering non-universal features 
associated with the finite band width $W$, it will be always assumed that all 
correlators involve energy differences much smaller than $W$. Final results 
are therefore expressed in zeroth order in $\Delta /W$, i.e. when the number 
of levels in the band tends to infinity. 

The paper is organized as follows: A derivation of the ballistic nonlinear 
$\sigma$-model is presented in section~II. The interpretation of the 
resulting functional integral will be discussed in section~III. There it is 
shown that regularization of the functional integral forces one to identify 
the low lying modes of the field theory as the Perron-Frobenius modes 
associated with irreversible classical dynamics. In section~IV we present two 
applications: the reduction of our model to RMT, and a calculation of the 
two-point DoS correlator beyond the universal regime. At the end of this 
section the range of validity of our approach is discussed.  
In section~V we generalize the model to incorporate non-semiclassical 
corrections such as weak scattering from $\delta$-correlated random impurities.
This establishes the relation between the ballistic $\sigma$-model and the 
conventional diffusive counterpart. A summary and discussion of the results 
is presented in section~VI.

\begin{figure}
  \begin{center}
 \leavevmode
    \epsfxsize=11cm 
\epsfbox{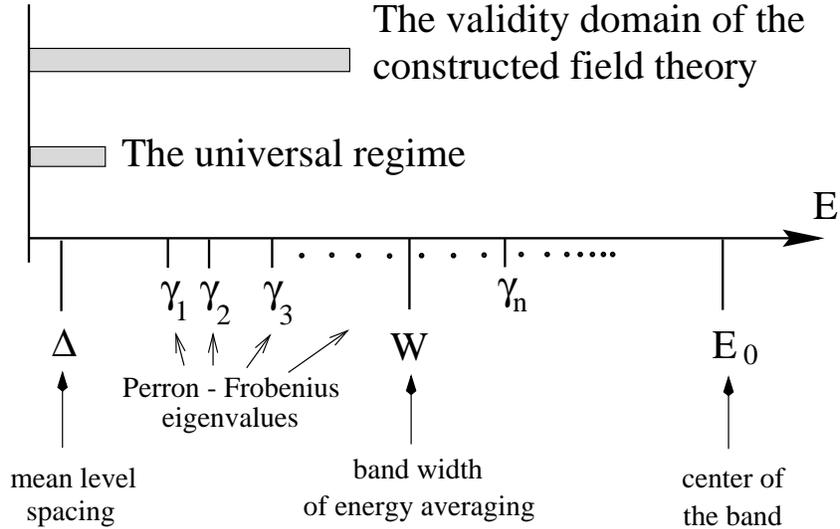}
 \end{center}
  \caption{A schematic diagram showing the relevant energy scales, as well as 
          the domains of validity of RMT and the field theory constructed in 
          this paper.}
  \label{fig:1}
\end{figure}

\section{The Nonlinear $\sigma$-Model} 
\label{sec:sigmamod}

To present the derivation of the effective field theory describing spectral 
correlations of quantum chaotic systems, we will focus on the problem of a 
single particle confined by an irregular potential described by 
the Hamiltonian
\begin{equation}
\hat{H}={\hat{\pp}^2\over 2m}+V(\hat{\pq}).
\end{equation}
The classical counterpart of the quantum Hamiltonian is assumed to be 
chaotic and to have no discrete symmetries. We confine attention to 
closed systems so that classical motion inhabits a finite region of 
the $2d$-dimensional phase space. We will
assume that all classical orbits are unstable and, in particular, 
exclude (KAM) systems where the phase space contains islands of regular 
motion. 

We will concentrate on statistical properties defined on an energy band 
of width $\win$ centered at an energy $E_0$. To discuss meaningful averages
it is necessary to assume that the average DoS, specified by the Weyl formula
\begin{equation} 
\langle \nu(E_0) \rangle  =
{1\over h^d}\int d\px\ \delta\left[E_0-H(\px)\right],\qquad
\px= (\pq,\pp) 
\label{weyl} 
\end{equation} 
is approximately constant within this interval. Taking as an example a 
particle in a random impurity potential, the accuracy of this approximation 
is of order $\win/E_0$, and can be made arbitrarily small by going into the 
semiclassical limit $E_0 \to \infty$. On the other hand, the bandwidth 
is assumed to be sufficiently large that the number of levels, $N=\nu(E_0) 
\win\gg 1$ can be employed as an expansion parameter --- final expressions 
will be expressed in the zeroth order approximation in $1/N$. Henceforth we 
will express energy in units of the mean level spacing, $\Delta=1/\langle
\nu(E_0)\rangle$ and denote such energies by $\epsilon =E/\Delta$. 
For simplicity, it is convenient to employ Gaussian averaging
\begin{equation} 
\langle \cdots \rangle_{\epsilon_0}=\int  
\frac{d\epsilon}{(2\pi N^2)^{1/2}} \exp\left[ 
-\frac{(\epsilon-\epsilon_0)^2}{2 N^2}\right] (\cdots ). 
\label{av} 
\end{equation}

A general $n$-point correlator of physical operators, such as the local or
global DoS or current densities, can be obtained from a generating function 
which depends on appropriate external sources. Here we focus on two-point 
correlators. Expressed as a field integral, 
the generating function for two-point correlators takes the form
\begin{eqnarray} 
{\cal Z}(\hat{J})&=&z\int D\Psi\ \exp\left[-\frac{i}{2}\int d \pq \Psi^\dagger 
(\pq) L\left(\hat{G}^{-1}(\epsilon)-\hat{J}\right)\Psi(\pq)\right],
\label{zdef} 
\end{eqnarray} 
where $\hat{G}^{-1}(\epsilon)=\epsilon-s^+\Lambda/2-\hat{H}$ denotes the 
matrix Green function with energy difference $s$ between retarded (R) and 
advanced (A) blocks, $\hat{J}$ represents the source, and the 
constant $z$ is included to enforce the correct normalisation.\footnote{ 
Energy averaging of the generating functional (\ref{zdef}) induces a quartic 
interaction of the form $(\Psi^\dagger L \Psi)^2$ among the supervector 
fields. The matrix $L$ serves as a metric tensor. In 
Ref.~\cite{Verbaarschot85} it was shown that the appropriate group of 
transformations preserving the interactions in the fermionic sector is 
compact while in the bosonic sector it must be chosen non-compact. This 
fixes the definition of $L$ (see Eq.~(\ref{L})). 
With this definition, the constant
\begin{equation}
z=\exp\left[{1\over 2}{\rm STr}_{\pq}\ln(\Lambda L)\right],
\end{equation}
accounts for the correct normalisation. For the definition of ${\rm STr}_{\pq}$
see discussion below Eq. (\ref{newhubstrat}).} Following 
Ref.~\cite{Altland93} we express the $8$-component superfields 
$\Psi$, which appear in Eq.~(\ref{zdef}), in the block notation
\begin{equation}
\label{Psi}
\Psi^p_{gd}=\left(\matrix{\Psi^A_{gd}\cr \Psi^R_{gd}\cr}\right)^p,\quad 
\Psi^p_{g=1,d}=\left(\matrix{\chi^p\cr {\chi^p}^*\cr}\right)_d,\quad
\Psi^p_{g=2,d}=\left(\matrix{S^p\cr {S^p}^*\cr}\right)_d,
\end{equation}
where superscript $p$ refers to retarded/advanced 
components, subscript $g$ refers to fermionic (F) components 
$\chi$ and bosonic (B) components $S$, 
and subscript $d$ refers to time-reversal (complex conjugated) 
components. The introduction of equal numbers of bosonic and fermionic 
fields is a standard trick which obviates the need to introduce replicas
and normalizes the generating function to unity. Matrices
\begin{equation}
\Lambda=\left(\matrix{1&0\cr0&-1\cr}\right)^p \otimes \openone_g \otimes 
\openone_d,\qquad 
k= \left(\matrix{1&0\cr0&-1\cr}\right)_g \otimes 
\openone_d ,\qquad 
\end{equation}
break the symmetry between the advanced/retarded and graded components 
respectively, and  we have 
chosen a convention which introduces the supermatrix 
\begin{equation}
\label{L}
L=\left(\matrix{\openone &0\cr0&k\cr}\right)^p \otimes \openone_d.
\end{equation}

The inclusion of complex conjugated fields effectively doubles the number of
fields and implies the 
relation 
\begin{equation}
\Psi^\dagger(\pq)=\Psi(\pq)^T C^T,
\label{psisym}
\end{equation}
where the operations of complex conjugation, and 
transposition of supervectors are defined following Efetov~\cite{Efetov83},
while 
\begin{equation}
\label{chargeconjug}
C=\openone^p \otimes \left(\matrix{-i\tau_2&0\cr0&\tau_1\cr}\right)_g,
\end{equation}
denotes the ``charge conjugation'' matrix, and
\begin{equation}
  \tau_0=\left(
  \begin{array}{cc}
    1 & 0 \\
    0 & 1
  \end{array}\right)_d, \qquad
  \label{Paulimat}
  \tau_1=\left(
  \begin{array}{cc}
    0 & 1 \\
    1 & 0
  \end{array}\right)_d, \qquad 
\tau_2=\left(
  \begin{array}{cc}
    0 & -i \\
    i & 0
  \end{array}\right)_d, \qquad 
\tau_3=\left(
  \begin{array}{cc}
    1 & 0 \\
    0 & -1
  \end{array}\right)_d
\end{equation}
represent Pauli matrices which act inside time-reversal blocks.\footnote{This 
convention differs slightly from that discussed by Ref.~\cite{Efetov83}.}

As an application, Eq.~(\ref{zdef}) can be used to represent the 
fluctuations in the two-point correlator of DoS by defining the source as
\begin{equation}
  \label{sourcedef}
  \hat{J}=J \delta(\pq-\pq') k \Lambda, 
\end{equation}
where $J$ is a constant. Then $R_2(s)$, defined in Eq.~(\ref{R2}), can be 
expressed as 
\begin{equation}
R_2(s)=-\frac{1}{16\pi^2}\frac{\partial ^2}{\partial J^2}
\Re \langle {\cal Z}(J) \rangle_{\epsilon_0} \Big|_{J=0}.
\end{equation}

If the energy difference $s$ is chosen to be much smaller than the width of
the energy band $N$, correlators become independent of $N$ and of the 
particular shape of the band (whether it is Gaussian or Lorentzian, etc). 
Performing the energy averaging (\ref{av}) of ${\cal Z}(J)$ we obtain
\begin{mathletters}
\begin{eqnarray} 
\left\langle {\cal Z}(J)\right\rangle_{\epsilon_0}&=& 
z \int D\Psi\exp\left[-\frac{i}{2}\int d \pq 
\Psi^\dagger(\pq)L\left(\hat{G}^{-1}(\epsilon_0)
-\hat{J} \right)\Psi(\pq)-S_{\rm int}[\Psi]\right],
\label{zaveraged}
\\ 
S_{\rm int}&=&{N^2\over 8} \left( \int d \pq \Psi^\dagger(\pq)L  
\Psi (\pq)\right)^2. 
\label{zaveragedb}
\end{eqnarray} 
\end{mathletters}
Therefore, in contrast to an impurity averaging, energy averaging induces a  
{\em nonlocal} interaction of $\Psi$. This represents an important departure
from the usual consideration of random Hamiltonians. 

The next step involves the decoupling of the interaction induced in the  
averaging by means of a Hubbard-Stratonovich transformation. This is achieved
with the introduction of $8\times 8$-component supermatrix fields 
$\hat{Q}(\pq_1,\pq_2)$ which are {\em non-local} in space. To define the 
correct decoupling it is crucial to identify those contributions to 
$S_{\rm int}$ which vary slowly in comparison with the wavelength. In the 
semi-classical analysis that follows we will show that the low lying degrees 
of freedom are described by matrices $\hat{Q}(\pq_1,\pq_2)$ which vary slowly 
with respect to the center-of-mass coordinate $(\pq_1+\pq_2)/2$. Anticipating 
this, it is convenient to switch to a momentum space representation of the 
interaction and explicitly separate such contributions
\begin{eqnarray}
  \label{slowint}
  S_{\rm int}={N^2\over 8}  \int d \pP \int_{ |\pp |<p_0} d \pp  
\left[ \Big( \Psi^\dagger(\pP )L  
\Psi (-\pP) \Big) \Big( \Psi^\dagger(\pP +\pp )L  
\Psi (-\pP -\pp) \Big) \right. \nonumber \\
\left. + \Big( \Psi^\dagger(\pP )L  
\Psi (-\pP) \Big) \Big(  \Psi^\dagger(-\pP -\pp )L  
\Psi (\pP +\pp) \Big) \right].
\end{eqnarray}
The characteristic momentum cut-off is defined such that $p_0 < \win/v$, 
where $v$ is the velocity of the particle. Using the charge conjugation 
symmetry of $\Psi$~(\ref{psisym}) it is straightforward to show that 
these terms give an equivalent contribution. 
Therefore, following Ref.~\cite{Efetov83} we decouple the 
interaction (\ref{zaveragedb}) as
\begin{equation} 
\label{newhubstrat} 
e^{-S_{\rm int}[\Psi]}=\int DQ
\exp\left[- \ {\rm STr}_\pq \left(\frac{\hat{Q}^2}{4} + {N \over 2} 
\Psi^\dagger L \hat{Q} \Psi \right)\right], 
\end{equation} 
where ${\rm STr}_\pq$ denotes the trace operation for supermatrices, 
${\rm STr}M={\rm Tr}M_{FF}-{\rm Tr}M_{BB}$, with a subscript $\pq$ used 
to denote a further extension of the trace to  include the coordinate 
integration. 
Eq.~(\ref{psisym}) implies that the dyadic product $A(\pq,\pq')=\Psi(\pq)
\otimes \Psi^\dagger(\pq')L$ obeys the symmetry property $A(\pq,\pq')= 
C^T L A^T(\pq',\pq) L C$. This induces the corresponding symmetry
\begin{equation}
  \label{chargeQ}
 \hat{Q}=C^T L\hat{Q}^TL C,
\end{equation}
where the transposition should be understood in the sense of an operator. 

Substituting Eq.~(\ref{newhubstrat}) into Eq.~(\ref{zaveraged}) and 
integrating over $\Psi$ we obtain the following expression for the 
averaged generating functional 
\begin{mathletters} 
\begin{eqnarray} 
\label{genfunc} 
\langle {\cal Z}(\hat{J})\rangle_{\epsilon_0}&=&\int DQ
\ \exp\left[-{ 1\over 4} {\rm STr}_\pq \hat{Q}^2+{1\over 2}{\rm STr}_\pq 
\ln \left( \hat{\cal G}^{-1}(\hat{Q})- {s^+\over 2}\Lambda 
-\hat{J} \right)\right],
\\ 
\label{efftheory} 
\hat{\cal G}^{-1}(\hat{Q})&=& \epsilon_0-{\hat H}- i N \hat{Q}.
\end{eqnarray} 
\end{mathletters} 

Thus far no approximations have been made. The next step is to identify the low
energy degrees of freedom and obtain an effective action. To do so, we will
employ a saddle-point approximation and find the matrix $\hat{Q}_0$ 
which minimizes the action in Eq.~(\ref{genfunc}). The effective field theory
is described by the expansion of the action in fluctuations of $\hat{Q}$ 
around the saddle-point. These fluctuations are strongly anisotropic and can 
be classified into massive and massless modes. The integral over the former 
can be  evaluated within the saddle-point approximation to leading order
in $1/N$ (see Appendix \ref{sec:massive}). The integral over the remaining 
massless modes, which arise from the underlying symmetry of the action 
(\ref{zaveraged}) must be evaluated exactly. The resulting field theory 
has the form of a nonlinear $\sigma$-model. 

\subsection{Saddle-point approximation and the $\sigma$-model}

Varying the action in Eq.~(\ref{genfunc}) with respect to $\hat{Q}$, and
neglecting the terms of order $s$ and $\hat{J}$, we find minima at 
$\hat{Q}_0$ which satisfy the equation 
\begin{equation} 
\label{saddlept} 
\hat{Q}_0\ \hat{\cal G}^{-1}(\hat{Q}_0) = -i N, 
\end{equation} 
where $\hat{Q}_0$ must be treated as an operator. The saddle-point 
solution which is diagonal in superspace is  given by
\begin{eqnarray} 
\label{spsolut}
\hat{Q}_0 &=& -i\frac{\epsilon_0-\hat{H} }{2 N}+
  \left[ 1 -\left( { \epsilon_0-\hat{H}  \over 2 N }
 \right)^2 \right] ^{1/2} \Lambda.
\end{eqnarray} 
Note that $N \hat{Q}_0$ plays the role of the self-energy 
in the average Green function ${\cal G}(\hat{Q}_0)$. 

The saddle-point solution in Eq.~(\ref{spsolut}) is not unique but is in fact
one member of a degenerate manifold of solutions. Their existence follows 
from the underlying symmetry of the action of Eq.~(\ref{zaveraged}). The 
interaction term $S_{\rm int}[\Psi]$ is invariant under the group of 
transformations $\Psi \to \hat{U}\Psi$ such that 
\begin{equation}
  \label{group}
  \hat{U}^\dagger L\hat{U}=L, 
\end{equation}
where $\hat{U}$ is an operator in Hilbert space. Terms that break the 
symmetry of the total action in Eq.~(\ref{zaveraged}) are $s\Lambda$, $\hat{J}
k\Lambda$ and the commutator $[\hat{H},\hat{U}]$. The invariance of the
relation $\Psi^\dagger =\Psi^T C^T$ under the transformation $\Psi \to 
\hat{U}\Psi$ induces an additional constraint on $\hat{U}$
\begin{equation}
  \label{Ucharge}
  \hat{U}^\dagger = C \hat{U}^T C^T.
\end{equation}
From Eq.~(\ref{newhubstrat}) it follows that these transformations 
induce  the following constraint on the Hubbard-Stratonovich field 
$\hat{Q}$: $\hat{Q} \to \hat{U}^{-1}\hat{Q}\hat{U}$. 

The saddle-point solution in Eq.~(\ref{spsolut}) is not invariant under this 
group of transformations. Therefore the low energy modes of the action are of 
the form $\hat{Q}=\hat{U}^{-1}\hat{Q}_0\hat{U}$. 
 However, not all of these transformations should be taken 
into account. The group of transformations 
(\ref{group}) contains a subgroup of matrices which commute with the 
Hamiltonian. The matrix $\hat{Q}$ remains diagonal in 
Hilbert space in the basis of eigenstates of the Hamiltonian. 
In Appendix \ref{sec:massive} we show that the massive mode integration 
gives rise to a suppression of the fluctuations of $\hat{Q}$ by the 
large parameter $N$. The only matrix $\hat{U}$ commuting 
with $\hat{H}$ which ``survives'' the $N\to \infty$ limit 
is the one proportional to the unit matrix in Hilbert space. 
Admitting matrices $\hat{Q}$ of such form into Eq.~(\ref{genfunc}) we 
obtain 
\begin{equation}
  \label{effactiondef}
  \langle Z(\hat{J})\rangle_{\epsilon_0}=\int DQ
\ \exp\left(-S_{\em eff}[{Q}]\right),
\end{equation}
where
\begin{eqnarray}
\label{effactionop} 
S_{\em eff}[\hat{Q}]&=& -\frac{1}{2}
{\rm STr}_\pq\ln\left[\hat{\cal G}^{-1}(\hat{Q})-{s^+\over 2}\Lambda
-\hat{J} \right]\nonumber \\ &=& -\frac{1}{2}
{\rm STr}_\pq\ln\left[\hat{\cal G}^{-1}(\hat{Q}_0)-\hat{U}\left(
{s^+\over 2}\Lambda +\hat{J}\right) \hat{U}^{-1}-\hat{U}[\hat{H},
\hat{U}^{-1}]\right]. 
\end{eqnarray} 
The last three terms under the logarithm in Eq.~(\ref{effactionop}) are  
small as compared to the first, and an expansion can be made in them. Each 
order in this 
expansion brings an additional power of $1/N$, and suggests the inclusion
of just the leading order term: 
\begin{eqnarray} 
\label{sigmamodelop} 
S_{\em eff}[\hat{Q}]&=&\frac{i}{2N} 
\ {\rm STr}_\pq \left[ \hat{Q} \left(\frac{s^+}{2} \Lambda 
+\hat{J}-\hat{U}^{-1}[\hat{H},\hat{U}]\right)\right]. 
\end{eqnarray}
This approximation is justified only if $s \ll N$ and the 
commutator $[\hat{H},\hat{U}]$ is not anomalously large. 
The validity of this approximation must be considered individually 
for each system. In section~\ref{sec:dis} we will discuss an example where 
this is not the case, and one has to keep the second order expansion 
of the logarithm in Eq.~(\ref{effactionop}).

\subsection{Semiclassical approximation}

In the limit $\epsilon_0\to \infty$, the configurations of the $Q$-matrix 
that contribute substantially to the functional integral in 
Eq.~(\ref{zaveraged}) can be described within the semiclassical approximation.
It is therefore convenient to re-express all operators in the Wigner  
representation. Given an operator $\hat{{\cal O}}$ as a set of 
matrix elements ${\cal O}(\pq_1,\pq_2)$  between two 
position states at $\pq_1$ and $\pq_2$, its Wigner 
representation is a function of the phase space variables 
$\px=(\pq,\pp)$ defined by
\begin{equation}
  \label{wigner}
  {\cal O}(\px) =\int 
d^dq'\ \exp (i \pp \cdot \pq'/\hbar)\ {\cal O}(\pq+\pq'/2,
\pq-\pq'/2).
\end{equation}
We will use the fact that, in the 
semiclassical limit, the Wigner transform of a product of operators
is equal the product of the Wigner transformed operators,  
$({\cal O}_1{\cal O}_2) (\px)\to {\cal O}_1(\px)
{\cal O}_2 (\px)$, where ${\cal O}_{1,2}(\px)$ are 
smooth slowly varying functions on the quantum scale~\cite{Moyal49}.
In this approximation Eq.~(\ref{group}) becomes 
\begin{equation}
  \label{groupwig}
   U^\dagger(\px)LU(\px)=L,
\end{equation}
and implies that the matrices $U(\px)$ belong to the pseudounitary 
supergroup $U(2,2/4)$.
Expressed in the Wigner representation, the constraint in Eq.~(\ref{Ucharge}), 
\begin{equation}
  \label{Uchargewign}
  U^*(\pq,\pp)=C U(\pq,-\pp)C^T,
\end{equation}
shows that the matrices $U(\px)$ at different $\px$ are not 
independent. This agrees with the findings of Ref.~\cite{Muzykantskii95b}.

The massless modes in the Wigner representation are generated by those
matrices $U(\px)$ which do not commute with $\Lambda$. Such matrices, denoted 
by $T(\px)$, belong to the coset space ${\bf H}={\bf G}/{\bf K}=U(2,2/4)/
\left[U(2/2)\times U(2/2)\right]$ and, as implied by Eq.~(\ref{Uchargewign}), 
satisfy the symmetry relation
\begin{equation}
  \label{Tcharge}
  T^*(\pq,\pp)=C T(\pq,-\pp)C^T. 
\end{equation}
As mentioned above, those matrices $\hat{U}$ which commute with the 
Hamiltonian are strongly suppressed by massive modes (see 
Appendix~\ref{sec:massive}). In the semiclassical limit, we therefore 
admit only those matrices $T(\px)$ which are independent of the energy. 
The massless modes are then given by
\begin{equation}
  \label{QmasslessT}
  Q(\px) = T^{-1}(\px_\parallel)Q_0(H)T(\px_\parallel),
\end{equation}
where $\px_\parallel$ denotes a phase space coordinate on the energy shell
$\epsilon_0= H(\px)$.

Substituting $\hat{T}$ for $\hat{U}$ in Eq.~(\ref{sigmamodelop}), 
and applying the semiclassical approximation in which the 
commutator with the Hamiltonian becomes the Liouville operator $\hat{\cal L}$,
\begin{equation}
[\hat{H},\hat{T}] \to - i\hbar \hat{\cal L} T(\px_\parallel)
 = - i\hbar \left\{ T(\px_\parallel) , H \right\},
\label{Liouville}
\end{equation}
where 
\begin{equation}
\left\{A, B \right\}=
\sum_i\left[\frac{\partial A}{\partial q_i}\frac{\partial B}{\partial p_i}
-\frac{\partial B}{\partial q_i}\frac{\partial A}{\partial p_i}\right]
\end{equation}
denotes the Poisson bracket of $A$ and $B$, we obtain 
\begin{eqnarray} 
\label{sigmamodelac} 
S_{\em eff}[Q]&=&\frac{i}{2 N} \int \frac{ d\px}{h^d} 
\ {\rm STr} \left[  Q(\px) \left(\frac{s^+}{2} \Lambda +\hat{J} 
+i\hbar T^{-1}\hat{\cal L} T\right)
\right].
\end{eqnarray} 
Since the only dependence on the coordinate $x_\perp \equiv H(\px)$ 
normal to the energy shell enters through $Q_0(H)$, given 
by Eq.~(\ref{spsolut}), the integral over this  
variable can be performed and yields a factor $\pi N \Lambda$. 
Introducing the notation
\begin{equation} 
{\cal Q}(\px_\parallel)\equiv\frac{1}{\pi N} \int d H\ T^{-1}(\px_\parallel) 
Q_0(H) T(\px_\parallel)=T^{-1}(\px_\parallel) 
\Lambda T(\px_\parallel),
\label{param} 
\end{equation} 
we obtain the final expression 
\begin{eqnarray} 
\label{sigmamodel} 
S_{\em eff}[{\cal Q}]&=&\frac{i\pi}{2}\int \frac{d \px_\parallel}{h^d} 
\ {\rm STr} \Big[ {\cal Q} \left({ s^+ \over 2} \Lambda +\hat{J}
+i\hbar T^{-1}  \hat{\cal L} T
\right)  \Big].
\end{eqnarray} 
Here and henceforth when the arguments of ${\cal Q}$ and $T$ are omitted 
they should be understood as functions of $\px_\parallel$.\footnote{ 
Note that, since we adopt the convention in which the DoS is equal to unity, 
the phase space coordinates are normalized as $\int d \px_\parallel /h^d =1$.} 

The matrix ${\cal Q} (\px_\parallel)$ introduced in Eq.~(\ref{param}) 
satisfies the nonlinear and symmetry constraints
\begin{equation}
  \label{Qconstraints}
  {\cal Q}(\px_\parallel)^2=1, \qquad {\cal Q}(\pq,\pp)=C^T L 
{\cal Q}^T(\pq,-\pp) L C .
\end{equation}

Naively, since Eq.~(\ref{sigmamodel}) is expressed through the matrices $T$ 
rather than through ${\cal Q}=T^{-1}\Lambda T$, it appears that it does not 
correspond to a $\sigma$-model. However, a general property of $\sigma$-models 
is the invariance of the action under gauge transformations 
$T \to R T$, where $R$ commutes with $\Lambda$. Using the fact that the
Liouvillian $\hat{\cal L}=\dot{\px}_\parallel \cdot \nabla_{\px_\parallel}$ 
(where $\dot{\px}_\parallel$ is the phase space velocity) is a first order 
differential operator it is straightforward to show that, under a gauge 
transformation, the change of the action (\ref{sigmamodel}) is given by 
\begin{eqnarray} 
\label{gaugechange} 
\delta S_{\em eff}[{\cal Q}]&=&- \frac{\pi \hbar}{2}\int 
\frac{d \px_\parallel}{h^d} 
\ {\rm STr} \Big[
 \Lambda R^{-1} \hat{\cal L} R 
 \Big] = - \frac{\pi \hbar}{2}\int 
\frac{d \px_\parallel}{h^d} 
\ {\rm STr} \Big[
 \Lambda  \hat{\cal L}\ln R 
 \Big]=0.
\end{eqnarray} 
To arrive at the last equality we used the fact that the flow 
in phase space is incompressible:
$\nabla_{\px_\parallel} \cdot \dot{\px}_\parallel =0$. 

The kinetic part of the action in Eq.~(\ref{sigmamodel}) 
is equivalent to that introduced by Muzykantskii and Khmel'nitskii 
\cite{Muzykantskii95} where it is written in the form of the 
Wess-Zumino-Witten action 
\begin{equation} 
S_{\rm WZW}[{\cal Q}]=\frac{\pi \hbar}{8}\int \frac{d \px_\parallel}{h^d} 
\ \int_0^1 du\ {\rm STr} \left(\tilde{\cal Q}\left[ 
 {\partial \widetilde{\cal Q}\over 
\partial u} ,  \hat{\cal L} \widetilde{\cal Q} \right]
\right),
\label{wzw} 
\end{equation} 
where $\widetilde{\cal Q}$ is a smooth function of the auxiliary variable $u$ 
and $\px_\parallel$ with the boundary conditions: 
$\widetilde{\cal Q}(\px_\parallel,1)={\cal Q}(\px_\parallel)$ and 
$\widetilde{\cal Q}(\px_\parallel,0)=\Lambda$. The equivalence of 
Eq.~(\ref{wzw}) with the kinetic part of the action in Eq.~(\ref{sigmamodel}) 
can be established straightforwardly by substituting $\widetilde{\cal Q}=
\widetilde{T}^{-1}\Lambda \widetilde{T}$ and manipulating the various 
terms using the identities $\widetilde{T}^{-1} \hat{{\cal L}} \widetilde{T}=
-( \hat{{\cal L}} \widetilde{T}^{-1})\widetilde{T}$ and
$\widetilde{T}^{-1} \partial_u \widetilde{T}=
-( \partial_u \widetilde{T}^{-1})\widetilde{T}$ which follow
from $\hat{{\cal L}}\left( \widetilde{T}^{-1}\widetilde{T}\right)= 
\partial_u \left( \widetilde{T}^{-1}\widetilde{T} \right) =0$. 
An integration by parts with respect to $\px_\parallel$ shows that the 
resulting integrand is a total derivative with respect to $u$. The 
integration over $u$, with the boundary conditions 
$\widetilde{T}(\px_\parallel,1)= T(\px_\parallel)$ and 
$\widetilde{T}(\px_\parallel,0)=\openone$, leads to
the kinetic part of the action in Eq.~(\ref{sigmamodel}).
 
\subsection{Range of validity of the $\sigma$-model}

To clarify the domain of applicability of the nonlinear $\sigma$-model
in Eq.~(\ref{sigmamodel}) let us review the main steps involved in its 
derivation. The construction of the effective generating functional in 
Eq.~(\ref{genfunc}) involved purely formal manipulations which involved 
no approximation. To proceed beyond this expression we invoked a saddle-point 
approximation in which the fluctuations of the massive modes were neglected. 
The parameter which controlled this approximation was the inverse bandwidth 
$1/N$. 

The second approximation involved the replacement of quantum mechanical 
commutators by the semiclassical Poisson bracket. Such an approximation is 
justified at high energies where the shortest length scale is set by the 
wavelength of the particle. Finally, in treating fluctuations of the 
massless modes around the saddle-point, we treat only the leading order
term in the expansion. Formally, if the commutator $[\hat{H},\hat{T}]$ is not 
anomalously large, this approximation is also justified by large $N$. Since 
characteristic configurations of $T$ are assumed semiclassical this 
assumption can be violated only  if $\hat{H}$ 
contains some non-semiclassical contributions. 

The validity of this semiclassical approximation is discussed in more 
detail later in section~\ref{sec:dis} when we return to consider 
scattering from quantum impurities and the relation of the 
ballistic $\sigma$-model to the conventional diffusive 
nonlinear $\sigma$-model.

The symmetry breaking terms in the action place additional constraints on the 
range of validity. The expansion around the saddle-point relies on 
characteristic frequencies (or energy scales arising from the Poisson bracket 
in the action (\ref{sigmamodel})) being much smaller than the bandwidth $N$. 

The derivation of the $\sigma$-model of Eq.~(\ref{sigmamodel}) relies 
solely on the presence of energy averaging which allows us to neglect 
the contribution of massive modes in the functional integral. 
Indeed, energy averaging is crucial in the ballistic limit
even in the presence of disorder. This was emphasized in the study of 
Altland and Gefen~\cite{AltlandGefen} of spectral statistics of ballistic 
metallic grains. There it was pointed out that ensemble averages of 
spectral correlators differ from averages performed over both ensemble 
and energy. In the semiclassical language of periodic orbit theory
this difference emerges from trajectories which are not scattered 
by impurities \cite{AgamFishman} and  give rise to ``clean features'' 
in the quantum spectrum.  The neglect of interference terms among 
different trajectories (namely the diagonal approximation) is allowed only
upon energy averaging over a wide band. Otherwise, the interference 
among these trajectories is substantial. 
In section~\ref{sec:dis} we show that without energy averaging 
only the diffusive $\sigma$-model of Efetov~\cite{Efetov83} can strictly be 
justified. In this case, the large parameter which suppresses the 
fluctuations of the massive modes is $(\tau \Delta)^{-1}$. 

As numerical or experimental studies of quantum chaos always involve finite 
statistics it is worth discussing how the results of such studies should be 
compared to the expressions above. Since the exact DoS consists of 
$\delta$-functions at the positions of energy levels, all spectral 
correlators based on finite statistics are inevitably singular. As the 
statistics are increased by extending the energy window $W$ over which 
averaging is performed, these correlators remain singular. The meaningful 
way to define spectral correlations is by introducing a smoothed 
DoS, $\rho_\gamma (\epsilon)=(\gamma /\pi) \sum_i 1/((\epsilon -\epsilon_i)^2+
\gamma^2)$. For any finite $\gamma$, the limit $N\to \infty$ exists and 
generates a ``smooth'' function. As an example, consider the two-point 
correlator of DoS for the unitary random matrix ensemble: To obtain the 
universal expression $R_2(s) = \delta(s) -\sin^2(\pi s)/(\pi s)^2$ it is 
possible to take the limit $\gamma \to 0$ only after the limit $N \to \infty$. 
If, however, one keeps $N$ finite while decreasing $\gamma$, the 
expression will approach the universal result only until $\gamma \approx 
1 /N$ after which it will start to deviate with increasing magnitude.
The expressions for spectral correlators that we are describing here 
should be understood in exactly the same way. They are asymptotic 
expressions corresponding to the limit of $N \to \infty $ taken before 
the limit $\gamma \to 0$. When one deals with experimental/numerical 
data, which necessarily involve finite statistics, one should always 
keep the level width finite such that $\gamma > 1/N$.

\section{Regularization of the Functional Integral and Irreversibility
of the classical dynamics}
\label{sec:reg}
 
In this section we discuss the question of ultraviolet divergences 
of the $\sigma$-model and show that the regularization procedure 
forces us to understand $\hat{\cal L}$ in Eq.~(\ref{sigmamodel}) 
as the generator of irreversible classical evolution.

The functional integral in Eq.~(\ref{effactiondef}) with the action 
(\ref{sigmamodel}) suffers from ultraviolet divergences and must be 
regularized. The ultraviolet divergence is not an artifact of the 
approximations employed in the derivation of Eq.~(\ref{sigmamodel}) and 
is present even in the original expression (\ref{zdef}) for the ratio of 
quantum spectral determinants ${\rm Det}(\epsilon -\hat{H})$. Although 
supersymmetry of Eq.~(\ref{zdef}) 
improves the ultraviolet properties of the functional integral, in higher 
dimensions it is not sufficient to make the expression converge. Therefore 
an ultraviolet regulator needs to be introduced. This regularization 
induces the corresponding regularization on the functional 
integral in Eq.~(\ref{effactiondef}). The kinetic part of the effective 
action (\ref{sigmamodel}) can be presented as ${\rm STr} [{\cal Q}  
T^{-1}\dot{\px}_\parallel \nabla_{\px_\parallel}T]$, where 
$\dot{\px}_\parallel$ is the classical phase space velocity. This 
action is only sensitive to the variations of the ${\cal Q}$-matrix along 
the classical trajectories. Therefore nothing prevents the ${\cal Q}$-field 
from fluctuating in the directions transverse to $\dot{\px}_\parallel$. 
It is these short scale fluctuations that ultimately lead to the divergence 
of the functional integral. The ultraviolet divergences are independent of 
the classical dynamics and of the shape of the constant energy surface and 
are unphysical. The diverging contribution to the functional integral 
therefore needs to be extracted by an appropriate regularization 
procedure. 

The problem of the ultraviolet regularization of functional integrals 
is well studied in field theory. One of the ways to regularize the functional 
integral is to introduce a term
\begin{equation} 
\delta S_R= m \int d \px_\parallel {\rm STr}(\nabla_{\px_\parallel} 
{\cal Q})^2,
\label{regulator} 
\end{equation} 
into the effective action (\ref{sigmamodel}). This term suppresses strong 
fluctuations of ${\cal Q}$ in the directions transverse to 
$\dot{\px}_\parallel$, and favors the physical functions 
${\cal Q}(\px_\parallel)$ which are smooth. Depending on the dimensionality of 
the phase space this may not be sufficient to make the integral convergent and 
additional regularization procedures should be invoked. To explore this issue 
in more detail we will consider the functional integral which arises from the 
lowest order perturbative expansion of the action.
 
In this case we can represent $T=1+\delta T$ and expand the action  
(\ref{sigmamodel}) with the regulator (\ref{regulator}) 
to second order in $\delta T$. We refer the reader to section~\ref{pert}
where this is discussed in greater detail. Here we only outline the conceptual 
steps which relate to the regularization. 

In the lowest order of perturbation theory, the resulting Gaussian functional 
integral generates simply the determinant of the elliptic operator 
$is-\hat{\cal L}_R=is- \hat{\cal L}-m \nabla_{\px_\parallel}^2 $ 
(see section~\ref{pert}). An operator is called elliptic is the
component of highest rank in derivatives is positive definite. The problem of 
regularization of the determinants of such operators is discussed in the 
literature (see, for example, Ref.~\cite{Schwarz}). One method involves the 
construction of a zeta function of the operator defined as 
\begin{equation} 
 \zeta (is-\hat{\cal L}_R |z) =\sum_i{1\over (is- \lambda_i)^z}= 
\frac{1}{\Gamma (z)}\int_0^\infty  t^{-z -1} dt {\rm  Tr} 
\exp\left[-\left(is - \hat{\cal L}_R \right)t \right].
 \label{operzeta} 
\end{equation} 
Here $\lambda_i$ denote the eigenvalues of $\hat{\cal L}_R$, and
we assume that $s$ is chosen such that the operator $is-\hat{\cal L}_R$ has 
no zero modes. Then the integral in the right hand side converges 
at the upper limit. At the lower limit $t \to 0$ it can diverge depending 
on the value of $z$. However, this divergence is ultraviolet in nature and  
can be removed by taking the integral at sufficiently large positive $z$.  
The expression can then be analytically continued to the rest of the 
complex plane. 

A regularized spectral determinant is expressed through the derivative of the 
zeta function (\ref{operzeta})
\begin{equation} 
 \left. \ln {\rm Det}(is-\hat{\cal L}_R )=-\zeta'
(is-\hat{\cal L}_R |z) \right|_{z=0}.
  \label{operdet} 
\end{equation} 
The regularized spectral determinant ${\rm Det}(is-\hat{\cal L}_R )$ is a 
function of $s$. It has zeroes at the positions of eigenvalues of 
$\hat{\cal L}_R$ and nowhere else on the complex plane. By taking the limit
one recovers the result which is independent of the regulator 
(\ref{regulator}) 
\begin{equation} 
{1\over Z(is)} = {\rm Det}(is-\hat{\cal L}) =\lim_{m\to 0}
{\rm Det}(is-\hat{\cal L}_R).
\label{detregless}
\end{equation}
This limit is very different for integrable and chaotic systems. In particular,
antihermiticity of $\hat{\cal L}$ in the limit $m\to 0$ suggests that the 
zeroes of the regularized determinant ${\rm Det}(is-\hat{\cal L})$ lie 
on the imaginary axis of $is$. However, for chaotic systems this is not the
case~\cite{Ruelle86,Pollicott90}. 

To understand the subtleties which arise when this limit is taken for 
nonintegrable systems, let us consider the purely classical evolution. Suppose
we form an initially non-uniform probability density distribution 
$\rho(\px_\parallel)$ in the phase space. The classical dynamics involves 
stretching along the unstable manifold and contraction along the stable one. 
Thus, any non-uniform initial distribution will evolve into a highly singular 
function along the stable manifold. The regularization term (\ref{regulator}) 
in the classical evolution can be ignored for short times but eventually, 
when contractions along the stable manifold make the phase space gradients 
sufficiently large, it becomes relevant.\footnote{An analogous situation 
arises in the theory of turbulence~\cite{Polyakov95}. In the inertial range,
viscosity can be neglected and turbulence can be considered as 
dissipationless. However at sufficiently small scales, velocity gradients 
become large and viscosity becomes relevant. In this picture the energy which 
is pumped into the system at large spatial scales is transfered without 
dissipation in the inertial range to smaller spatial scales and is eventually
absorbed at microscopic scales determined by viscosity. The latter can be 
viewed as an ultraviolet regulator, which is eventually set to zero but has a 
finite effect on the velocity correlators since it is necessary to produce a 
stationary solution.} Therefore the limits time-to-infinity and $m\to 0$ do 
not commute. To find the spectrum one has to take the time-to-infinity limit 
first and then set the regulator to zero. In this limit the eigenvalues 
$\gamma_\mu$ of $\hat{\cal L}$ have finite real parts corresponding to 
relaxation rates into the equilibrium distribution. These physical 
eigenvalues which reflect intrinsic irreversible properties of the {\em 
purely classical dynamics} are known as Ruelle resonances or the 
Perron-Frobenius spectrum~\cite{Ruelle86,Pollicott90}.  

Thus, the necessity to regularize the functional integral of the 
$\sigma$-model forces us to understand $\hat{\cal L}$ as 
the classical evolution operator which corresponds to the 
{\em irreversible} classical dynamics. 

There are several ways of calculating the Perron-Frobenius spectrum other 
than diagonalizing ${\cal L}_R$ and taking the ``zero noise limit'', 
$m\to 0$~\cite{Gaspard95}. These employ, for instance, symbolic 
dynamics~\cite{Ruelle78}, course graining of the flow dynamics in phase 
space~\cite{Nicolis}, and analytic continuation~\cite{Hasegawa92}. 
An exact formal expression for the dynamical zeta function 
$1/Z (z)$, which should be understood as a regularized product 
$\prod_{\mu} (z-\gamma_{\mu})$, is given in terms of the
classical periodic orbits of the system. For two-dimensional systems
it is of the form~\cite{Cvitanovic}
\begin{eqnarray}
{1\over Z(z)}= \prod_p \prod_{k=0}^{\infty}\left(1-
\frac{e^{zT_p}}{|\Lambda_p| \Lambda_p^k}\right)^{k+1},
\label{dynamical} 
\end {eqnarray}
where $T_p$ is the period of the $p$-th primitive orbit and $\Lambda$ is the 
eigenvalue of the Monodromy matrix with absolute value larger than one. 
(The Monodromy matrix is the linearized map on the Poincare surface of 
section in the vicinity of the orbit.) In its present form, $1/Z(z)$ cannot 
be used to determine the eigenvalues $\gamma_{\mu}$. For this purpose a 
re-summed formula is required. It can be obtained by expanding the infinite 
product over the periodic orbits and ordering the various terms in a way 
that leads to maximal cancellation among them. This method, known as the 
cycle expansion~\cite{Artuso90}, exploits the property that the dynamics of 
chaotic systems in phase space is coded by a skeleton of a small number of 
periodic orbits called fundamental orbits. In this sense, long periodic 
orbits may be viewed as linear combinations of the fundamental orbits.

To summarise the main conclusion of this section, it was shown that the 
regularization procedure which properly defines the functional integration 
forces one to understand the low lying degrees of freedom of the action 
(\ref{sigmamodel}) as the Perron-Frobenius relaxational modes of the 
classical counterpart. We emphasize that these modes represent purely 
classical characteristics of the system independent of the regularization 
procedure. Indeed the corresponding spectral determinant 
$\det(z - \hat{\cal L})$ has an exact representation in terms of the
classical periodic orbits of the system (\ref{dynamical}).

\section{Applications}

To interpret the findings of the previous sections we will apply the 
generalized nonlinear $\sigma$-model to the regime of long-time or low 
energy scales. This will establish a firm connection of level statistics 
with RMT. Corrections to RMT will be studied within the framework of a 
perturbation theory involving the modes of the Perron-Frobenius operator. 
These results indicate a close correspondence between spectral correlations
of the classical and quantum operators which we discuss. 

\subsection{Random Matrix Theory}

It is widely believed that the statistical quantum properties of systems with 
{\em few} degrees of freedom can be described, at least over some range of 
energy scales, by RMT \cite{Casati80,Bohigas84,Berry85}. To interpret this, 
various approaches have been developed largely along two parallel lines
discussed in the introduction. The first approach concerned the study of 
ensembles of random systems such as disordered metallic 
grains~\cite{Eliashberg65,Efetov83,Altshuler86}. Randomness in this case is 
introduced on the level of the Hamiltonian itself usually as a consequence of 
some impurity configuration. The second approach involves the study of 
non-stochastic systems which are chaotic in their classical limit such as the 
Sinai or the stadium billiards~\cite{Bohigas84}. In this case ``randomness'' 
is generated by the underlying 
deterministic classical dynamics itself. Nevertheless, it has been 
conjectured~\cite{Bohigas84} that spectral fluctuations of strongly chaotic
quantum systems are described by level statistics of random matrix ensembles.

Despite being supported by extensive numerical studies, the origin of the 
success of RMT as well as its domain of validity are still not completely 
resolved. Below we will show that, in the semiclassical limit, this
conjecture is indeed valid for chaotic systems without any discrete
symmetries, and which are characterized by an exponential decay of classical 
correlation functions in time.

If we define by $\{ \gamma_n \}$ the set of eigenvalues of the 
Perron-Frobenius operator $\hat{\cal L}$, then the lowest eigenvalue 
in ergodic systems is $\gamma_0=0$. This eigenvalue, associated with the 
invariant density on the energy shell, is non-degenerate and manifests the 
conservation of probability density. Any initial density distribution 
eventually relaxes to the state associated with $\gamma_0$. If, in addition, 
this relaxation is exponential in time, then the Perron-Frobenius spectrum 
has a gap associated with the slowest decay rate. Thus, for the first non-zero 
eigenvalue $\gamma_1$, we have $\gamma_1'\equiv \Re(\gamma_1)> 0$. This 
gap sets the ergodic time scale, $\tau_c=1/\gamma_1'$ over which the 
classical dynamics relaxes to equilibrium. In the case of disordered metallic 
grains, it coincides with the Thouless time, while in ballistic systems or 
billiards it is of order of the time of flight across the system.

In the limit $s\ll \gamma_1'$, or equivalently at times
which are much longer than $\tau_c$, the dominant contribution
to Eq.~(\ref{effactiondef}) with the effective action of 
Eq.~(\ref{sigmamodel}) arises from the ergodic classical distribution,
the zero-mode $\hat{\cal L}T_0=0$. Taking only this contribution, the 
functional integral (\ref{effactiondef}) becomes definite. Previous studies 
have demonstrated the equivalence of the zero-mode action with Wigner-Dyson
level statistics of RMT~\cite{Efetov83,Verbaarschot85}. For example, taking
the source as given by Eq.~(\ref{sourcedef}), we obtain
\begin{equation}
\langle {\cal Z}(J) \rangle_{\epsilon_0} = \int
d Q_0  \exp \left[ -i\frac{\pi}{4} \mbox{STr}
\left(\left[ s^+ +2Jk\right]\Lambda {Q}_0\right) \right],
\end{equation}
where ${Q}_0= {T}_0^{-1} \Lambda {T}_0$. This expression coincides with that
obtained in Ref.~\cite{Efetov83} and reproduces Wigner-Dyson level 
correlations. We therefore conclude that the quantum statistics of chaotic 
systems having no discrete symmetries and with exponential classical 
relaxation are described by RMT at energies smaller than $\gamma_1'$.

The RMT description is expected to hold even for certain chaotic
systems where the Perron-Frobenius spectrum is gapless~\cite{Bohigas84}. 
Examples include the stadium or Sinai billiards where classical 
correlation functions decay algebraically in time~\cite{Bunimovich}. In this
case, the resolvent $(z-\hat{\cal L})^{-1}$ is expected to have cuts which 
reach the $\Im z$ axis. Nevertheless, we expect the RMT description to hold 
whenever the spectral weight of the resolvent inside the strip $0\le \Re z 
\le 1$ (which, however, excludes the pole at the origin) is much smaller than 
unity. 

\subsection{The two-point correlation function: Beyond universality}
\label{pert} 

In this section we will make use of the $\sigma$-model to examine how 
corrections to RMT appear at larger energy scales. Again, focusing on the 
two-point DoS correlator, the generating function leads to the expression
\begin{equation} 
R_2(s)=\frac{1}{64}\Re\int {\cal D}Q 
\left( \int d \px_\parallel {\rm STr} 
[\Lambda k Q(\px_\parallel)]\right)^2
\exp\left[-S_{\em eff}(s)\right],
\label{r2gen} 
\end{equation} 
where  
\begin{eqnarray}
  \label{ZR2}
  S_{\em eff}(s)&=& 
 {\pi\over 2}\int d\px_\parallel
 {\rm STr} \left[i\frac{s^+}{2} \Lambda Q +QT^{-1}\hat{\cal L} T\right].
\end{eqnarray}
Although straightforward, the perturbative expansion is somewhat technical,
and here we present only the results of the detailed calculation described 
in Appendix~\ref{perturbation}.

In the limit of high frequencies $s \gg 1$, 
the two-point correlator takes the asymptotic form 
\begin{equation}
R_2(s)=R_P(s)+R_{NP}(s),  
\label{rpertosc}
\end{equation}
where both the nonperturbative term $R_{NP}(s)$ as well as the perturbative 
one $R_P(s)$ are expressed through the classical spectral determinant 
${\cal D}(s)$ as
\begin{equation}
R_{NP}(s)= \frac{\cos(2\pi s)}{2\pi^4} {\cal D}^2(s), \quad  
R_P(s)= -\frac{1}{ \pi^2} \frac{\partial^2}{\partial s^2}
\ln [{\cal D}(s)].
\label{perturbative}
\end{equation} 
The determinant ${\cal D}(s)$, regularized according to the procedure outlined
in section~\ref{sec:reg}, is expressed in terms of determinants of the 
Perron-Frobenius operator 
\begin{eqnarray}
  \label{ZR2pert}
  {\cal D}(s)&=&\Re\frac{ {\rm Det}'( \hat{ \cal L})^2}
{{\rm Det}\left[(is   - \hat{\cal L}) (-is  - \hat{\cal L})\right]},
\end{eqnarray}
where the prime indicates that the zero eigenvalue should be excluded from
the determinant. ${\cal D}(s)$ can be expressed in terms of the eigenvalues 
$\gamma_{\mu}$ of $\hat{\cal L}$, the Ruelle resonances, as
\begin{equation}
{\cal D}(s)=\prod_\mu {A^2(\gamma_{\mu})\over \left(\gamma_{\mu}^2+ s^2
\right)^2},
\label{spectraldet}
\end{equation} 
where $A(\gamma_{\mu})=\gamma_{\mu}^2$ for $\gamma_{\mu}\neq 0$ and 
$A(\gamma_{0}=0)= 1$. Note that, if the product in Eq.~(\ref{spectraldet}) is 
formally divergent, ${\cal D}(s)$ should be understood as the regularized 
determinant (\ref{detregless}). 

These results agree with those conjectured in Ref.~\cite{Agam95} and compare
with the perturbative expressions previously found for weakly disordered 
metals~\cite{Altshuler86,Andreev95,Jona-lasinio96} when the eigenvalues 
of the Liouville operator are identified by the eigenvalues of the 
diffusion operator. 

\section{Beyond the Semiclassical Approximation}
\label{diffsigma}

The derivation of the nonlinear $\sigma$-model in Eq.~(\ref{sigmamodel}) 
relied on the use of the semiclassical approximation. However, often we are 
concerned with quantum chaotic systems which can not be treated 
straightforwardly within the framework of semiclassics. A familiar example 
involves the quantum mechanical scattering of particles from a weak random 
impurity potential. In such cases, a formal justification of the ballistic 
nonlinear $\sigma$-model in Eq.~(\ref{sigmamodel}) does not seem possible. 
However, if the quantum Hamiltonian can be resolved into a part that can be 
treated within semiclassics and a part which can not, when the latter is 
small, a perturbation treatment may still be possible. 

Consider a general Hamiltonian $\hat{H}$ with matrix elements
\begin{equation}
\hat{H}=\hat{H}_{\rm cl}+\hat{H}_{\rm qu},
\end{equation}
where $\hat{H}_{\rm cl}$ represents the contribution which can be treated 
within a semiclassical approximation, and $\hat{H}_{\rm qu}$ determines the 
part which can not.

If the matrix elements of $\hat{H}_{\rm qu}$ are small as compared 
to the band width $N$ (a more precise criterion for a specific operator 
$\hat{H}_{\rm qu}$ is formulated below) their effect can be treated within 
the $\sigma$-model approach. In this case the saddle-point is governed by 
$\hat{H}_{\rm cl}$, and we can use Eq.~(\ref{spsolut}) with $\hat{H}$ 
replaced by $\hat{H}_{\rm cl}$. The contribution of $\hat{H}_{\rm qu}$ to the 
effective action can be found by expanding Eq.~(\ref{effactionop}),
\begin{eqnarray}
\label{effactionop2} 
S_{\em eff}[\hat{Q}]&=& -\frac{1}{2}
{\rm STr}_\pq\ln\left[\hat{\cal G}^{-1}(\hat{Q}_0)-\hat{H}_{\rm qu}-
\hat{U}\left({s^+\over 2}\Lambda+\hat{J}\right)\hat{U}^{-1}-\hat{U}[\hat{H},
\hat{U}^{-1}]\right], 
\end{eqnarray} 
where the supermatrix Green function involves only $\hat{H}_{\rm cl}$. 
Expanding to second order in $\hat{H}_{\rm qu}$ we obtain
\begin{eqnarray} 
\label{sigmamodelop2} 
S_{\em eff}[\hat{Q}]&=&\frac{1}{2N} 
\ {\rm STr}_\pq \left[ i\hat{Q} \left(\frac{s^+}{2} \Lambda 
+\hat{J}+\hat{H}_{\rm qu}-\hat{U}^{-1}[\hat{H}_{\rm cl},\hat{U}]\right)
+{1\over 2 N}\left(\hat{Q} \hat{H}_{\rm qu}\right)^2\right]. 
\end{eqnarray}
Finally, representing the $Q$-matrices in the Wigner representation,
the second order correction to the action takes the form
\begin{eqnarray}
-{1\over 4 N^2}{\rm STr}_{\pq}\left(\hat{Q} \hat{H}_{\rm qu}\right)^2
&=&-{1\over 4 N^2} \int \prod_{i=1}^4 dq_i \prod_{i=1}^2 {dp_i\over h^d}
e^{-i\pp_1(\pq_1-\pq_2)/2\hbar-i\pp_2(\pq_3-\pq_4)/2\hbar} \nonumber
\\ & &\times H_{\rm qu}(\pq_2,\pq_3) H_{\rm qu}(\pq_4,\pq_1)
{\rm STr}\left[Q\left(\pp_1,(\pq_1+\pq_2)/2\right) 
Q\left(\pp_2,(\pq_3+\pq_4)/2\right)\right],
\label{secondo}
\end{eqnarray}
where $H_{\rm qu}(\pq ,\pq ') = 
\langle \pq |\hat{H}_{\rm qu}|\pq '\rangle $.

\subsection{Random impurities and the restoration of the diffusive 
nonlinear $\sigma$-model} 
\label{sec:dis} 

To illustrate these ideas, let us consider the physical example involving a
particle moving in a background of weakly scattering impurities. If the 
${\cal Q}$ matrices vary on a scale that is long as compared to the 
scattering length $\ell=v \tau$, the particle dynamics becomes diffusive and 
we should recover the supersymmetric nonlinear $\sigma$-model obtained by 
Efetov~\cite{Efetov83}. In the opposite limit, the impurities generate a new 
term in the action which takes the form of a collision integral. 

The problem of dilute scattering impurities in an otherwise ballistic system 
has been discussed previously. A description within the framework of 
diagrammatic perturbation theory was investigated by Altland and 
Gefen~\cite{AltlandGefen}. More recently, in an important development, 
Muzykantskii and Khmelnitskii~\cite{Muzykantskii95} introduced an 
effective field theory to extend the diffusive $\sigma$-model 
into the ballistic regime. 

For simplicity, let us consider a (dimensionless) $\delta$-correlated white 
noise Gaussian random potential
\begin{equation}
H_{\rm qu}(\pq,\pq^\prime)=V(\pq)\delta^d(\pq-\pq^\prime),
\end{equation}
with a second moment defined by the mean free time $\tau$, 
\begin{equation} 
\left\langle \delta V(\pq)\right\rangle_V=0 ,\qquad 
\left\langle \delta V(\pq) \delta V(\pq^\prime)\right\rangle_V=  
{\hbar\over 2\pi \nu \tau \Delta ^2} \delta^d(\pq-\pq^\prime).
\end{equation}
Here $\langle \cdots \rangle_V$ denotes the ensemble average over the 
random potential, $\nu =1/\Delta \Omega$ represents the average local DoS, and 
$\Omega$ is the volume of the system. 

In this case, the expansion of the action around the saddle-point of the 
Hamiltonian $\hat{H}_{\rm cl}$ is justified in the limit 
$\hbar /\tau\ll N \Delta$. The same condition allows the truncation 
of the perturbation series at second order. Once again, performing 
the energy integration (\ref{param}) and using the fact that 
$d \px_\parallel /h^d = d \pq d \pp_\parallel /4 \pi p_F^2 \Omega$, where 
$\pp_\parallel$ is momentum on the constant energy shell ($|\pp_\parallel| =
p_F$), we obtain the effective action
\begin{eqnarray} 
\label{sigmamodel2} 
S_{\em eff}[{\cal Q}]&=&\frac{i\pi \nu}{2}\int 
\frac{d \pq d \pp_\parallel}{4\pi p_F^2 } 
\ {\rm Str} \left[\left(\Delta \left[{ s^+ \over 2} \Lambda +\hat{J}\right] 
-i\hbar T^{-1}\left\{ H_{\rm cl},T\right\}\right) {\cal Q} \right]\nonumber 
\\ &+ &
{\pi \nu \hbar \over 8\tau} \int 
\frac{d \pq d \pp_\parallel d \pp^\prime_\parallel}{(4\pi p_F^2)^2}
{\rm STr}\left[{\cal Q}(\pq, \pp_\parallel){\cal Q}(\pq, 
\pp^\prime_\parallel)\right].
\end{eqnarray} 
Although this action is precisely of the form of that introduced in 
Ref.~\cite{Muzykantskii95}, its derivation and the domain of validity seems 
far removed from that proposed in this earlier work. The $\sigma$-model 
description of the ballistic regime holds only if the frequencies of interest 
(or, equivalently the characteristic energies of the gradient terms) 
are small as compared with the width of the band 
$N \Delta$. In the absence of energy averaging the range of validity of this
description is restricted to the diffusive regime, where it coincides with the 
diffusive $\sigma$-model~\cite{Muzykantskii95}. At higher energies the 
massive modes have to be taken into account, and the $\sigma$-model 
description breaks down. The distinction drawn by energy averaging has been 
emphasized by Altland and Gefen~\cite{AltlandGefen}. Physically the 
difference arises from those orbits whose period is shorter than $\tau$ but 
longer than the inverse band width $( N\Delta)^{-1}$. Technically the energy 
averaging suppresses the massive mode fluctuations and facilitates 
the $\sigma$-model description.

To establish the relation between the ballistic $\sigma$-model and the 
conventional diffusive counterpart we follow Ref.~\cite{Muzykantskii95}. 
Let us suppose that the classical component of the Hamiltonian 
corresponds to free propagation. Anticipating a rapid relaxation of 
the momentum dependent degrees of freedom of ${\cal Q}$ on the energy 
shell, and a slow variation of the spatial modes, 
we introduce a parametrisation which involves the moment expansion
\begin{equation}
T(\px)=T_K(\pq) T_0(\pq) ,\qquad T_K(\pq)=\exp[i \pn(\pq) \cdot 
\pK(\pq)],
\end{equation}
where $\pn=\pp/|\pp|$ and, without any loss of generality, we choose 
$\Lambda \pK + \pK \Lambda =0$. To enforce the nonlinear 
constraint in the most convenient way, we have adopted a parametrisation 
which departs from that discussed in Ref.~\cite{Muzykantskii95}.

Expanding the action to second order in $\pK$ and performing integrals
over $\pn$ we obtain
\begin{equation}
S_{\rm eff}=\frac{ \pi \nu}{2}\int d\pq {\rm STr}
\ \left[i \Delta \left(\frac{s^+}{2}\Lambda + \hat{J} \right) Q 
-\frac{2 i \hbar  v_f}{3} \pK\cdot (\nabla T_0) 
T^{-1}_0 \Lambda -{\hbar \over 3 \tau} \pK^2
\right],
\end{equation}
where $Q(\pq)=T^{-1}_0(\pq)\Lambda T_0(\pq)$. Performing the Gaussian 
integration over $\pK$ we obtain the effective action
\begin{equation}
S_{\rm eff}={\pi \nu\over 8}\int d\pq {\rm STr}\ \left[\hbar D(\nabla Q)^2+
i2\Delta \left(s^+\Lambda + 2\hat{J} \right) Q\right],
\end{equation}
which coincides with that of the conventional diffusive 
$\sigma$-model~\cite{Efetov83}.

\section{Discussion}

In conclusion, we have shown that the quantum statistical properties of 
chaotic systems are described by a functional supersymmetric nonlinear 
$\sigma$-model with an effective action given by Eq.~(\ref{sigmamodel}). 
This result was obtained by employing energy averaging as opposed to ensemble 
averaging previously used for disordered metallic grains. The low lying 
degrees of freedom of the action (\ref{sigmamodel}) were identified as the 
Perron-Frobenius eigenmodes of the underlying classical dynamics. Thus, 
statistical characteristics of the quantum mechanical system in the 
semiclassical limit are determined by the symmetries of the system and 
properties of the Perron-Frobenius operator. In particular, a universal 
behaviour described by RMT is expected whenever the system has no symmetries 
and a gap exists in the Perron-Frobenius spectrum.

Our approach, however, assumes no systematic degeneracies of the actions 
of  the classical orbits of the system other than those associated with 
the known
discrete symmetries of the Hamiltonian. To emphasize this point, consider
the return probability to a given point. In the semiclassical limit it is 
given by a double sum over classical returning trajectories $\sum_{ij} A_i 
A^*_j$, where $A_i$ denotes the probability amplitude associated with the 
$i$-th path. If the corresponding actions are much larger than $\hbar$, the 
return probability reduces to a sum over probabilities, $\eta \sum_i 
|A_i|^2$. The factor $\eta$ is an integer which accounts for
exact degeneracies in the orbits of the action, which arise, for example,
from the existence of time reversal or reflection symmetries. When such a 
degeneracy is characterized by the existence of a discrete symmetry, it can, 
in principle, be incorporated into the $\sigma$-model. However, there are 
systems in which degeneracies of the actions cannot be characterized 
by simple discrete symmetries. Examples include the arithmetic billiards on 
surfaces of constant negative curvature~\cite{Bogomolny92}. 
The actions of the periodic orbits of these systems become 
exponentially degenerate as their length increases. This is a result of hidden 
symmetries which originate from number theoretic properties of these 
billiards. Indeed, despite showing a gap in the corresponding Perron-Frobenius 
spectrum, arithmetic billiards to not exhibit random matrix behaviour.
 
The $\sigma$-model derived here has a wide domain of validity which
goes well beyond the RMT results. Using perturbation theory the two-point 
DoS correlation function (\ref{R2}) was calculated. The result 
(shown in Eqs.~(\ref{rpertosc}) and (\ref{perturbative})) is expressed in 
terms of the spectral determinants of the Perron-Frobenius spectrum. Similar 
results were obtained recently by Bogomolny and Keating~\cite{Bogomolny96}. 
However, their results differ from ours by terms which are related to high 
repetitions of the same periodic orbit. The two results therefore clearly 
coincide in the limit where all orbits are highly unstable. At this stage 
the source of discrepancy is not understood. Whether it is related to 
corrections to the leading order of perturbation theory (Eqs. (\ref{rpertosc}) 
and (\ref{perturbative})) or to the nature of uncontrolled approximations
used by Bogomolny and Keating~\cite{Bogomolny96} remains unclear. It is 
appropriate, however, to mention that the $\sigma$-model functional integral 
with the action (\ref{sigmamodel}) can be solved exactly for the 
one-dimensional harmonic oscillator~\cite{Zirnbauerunpublished} and gives the 
correct result. This result can also be obtained by use of the perturbation 
theory described in section~\ref{pert}. Such a calculation 
shows that it is necessary to take into account the contributions of both 
stationary points of the action to obtain the exact result. The diagonal 
approximation of periodic orbit theory, on the other hand, also reproduces 
the exact result. Since all the periodic orbits of this system are 
repetitions of one primitive orbit, this result suggests that the diagonal
approximation commonly employed in periodic orbit theory, does {\it not} 
coincide precisely with the diagrammatic perturbation theory, as 
is commonly assumed.

Many of the results previously obtained for disordered systems concerning 
statistical properties of  wavefunctions and spectra depend only on the 
spectral properties of the diffusion operator in a given 
system~\cite{Kravtsov94,Fyodorov95}. These can be generalized 
straightforwardly to the case of chaotic systems by substituting the spectrum 
of the diffusion operator by the Perron-Frobenius spectrum.

The field theoretic approach described in this paper offers a systematic way 
of studying a variety of issues. These include; (i) the transition 
between the orthogonal and the unitary ensembles in ballistic systems; (ii) the
effects of discrete symmetries on spectral statistics in systems exhibiting
hard chaos; and (iii) weak localization effects in ballistic 
systems~\cite{Aleiner96}. 

We are grateful to K.~B.~Efetov, D.~E.~Khmel'nitskii, A.~I.~Larkin, 
I.~V.~Lerner, B.~A.~Muzykantskii, A.~M.~Polyakov, D.~Ruelle, Ya.~G.~Sinai, 
N.~Taniguchi, and M.~R.~Zirnbauer for stimulating discussions. This work was 
supported in part by JSEP No. DAAL 03-89-0001, and by the National Science 
Foundation under Grant Nos. DMR 92-14480 and PHY94-07194. We acknowledge the 
hospitality of the ITP in Santa Barbara where part of this work was 
performed. O.~A.~also acknowledges the support of the Rothschild Fellowship.

\appendix

\section{Saddle-Point Approximation: Identifying the Massive Modes} 
\label{sec:massive} 
 
In this section we examine fluctuations around the solution (\ref{spsolut}) 
of the saddle-point Eq.~(\ref{saddlept}) to identify the massive modes in the 
effective theory of Eq.~(\ref{efftheory}). Although we focus our remarks on
the orthogonal case studied in this paper, the general conclusions of this 
appendix hold for all ensembles.

To identify the massive modes it is convenient to work in the eigenbasis 
$\{ \varphi_n \}$ of the quantum Hamiltonian, where Eq.~(\ref{spsolut}) 
takes the form 
\begin{equation}
  \label{spsolutdiag}
  \left[Q_0\right]_{\mu \nu}=\delta_{\mu \nu}Q_\nu=
\delta_{\mu \nu}\left( {\epsilon_0-\epsilon_\mu
\over 2 N}+ \left[1 -\left( {\epsilon_0-\epsilon_\mu
\over 2 N }\right)^2 \right]^{1/2}\Lambda\right).
\end{equation}
 
Massive modes appear as fluctuations $\delta Q$ that commute with $Q_0$ in  
superspace. Expanding the action in Eq.~(\ref{efftheory}) around the  
saddle-point to second order in $\delta Q$ and neglecting $s$ (and using the
fact that $Q_0$ is diagonal in the Hilbert space indices) we obtain 
\begin{equation} 
\label{fluct} 
\delta S_2= -\frac{1}{2} \sum_{\mu \nu}{\rm Str}\ \left(Q_\mu \delta 
Q_{\mu \nu} Q_\nu \delta Q_{\nu \mu}+ \delta Q_{\mu \nu} \delta 
Q_{\nu \mu}\right). 
\end{equation} 
The mass of these modes is not apparently large but is of order one. 
However, their contribution to the two-point DoS correlator is given by
\begin{equation}
  \label{R2massive}
  R_{2, {\rm massive}}= 
-\frac{1}{(4\pi  N)^2} \sum_{\mu}\left\langle {\rm STr}(\Lambda k \delta 
Q_{\mu\mu})\ {\rm STr}(\Lambda k \delta Q_{\mu \mu})\right\rangle_Q 
\propto {1 \over N} ,
\end{equation}
where $\langle \cdots \rangle_Q$ denotes the average over supermatrices
$Q$ with respect to the action in Eq.~(\ref{genfunc}). 
This contribution vanishes in the $N \to \infty $ limit.

If we consider the contribution of the massive modes to a correlator of 
local observables such as the local DoS $\nu(\pq)$ we find
\begin{eqnarray} 
\label{nulocmassive}
\left\langle \delta \nu(\pq_1) \delta \nu(\pq_2)\right\rangle_{\rm massive}&=& 
-\frac{1}{(4\pi  N)^2} \sum_{\mu \nu}\varphi_\mu(\pq_1)
\varphi_\nu^*(\pq_1) \varphi^*_\mu(\pq_2)\varphi_\nu(\pq_2) 
\left\langle {\rm STr}(\Lambda k \delta Q_{\mu\nu})\ {\rm STr}(\Lambda k 
\delta Q_{\nu \mu})\right\rangle_Q \nonumber \\
& \propto & \frac{1}{  N^2} \sum_{\mu}\varphi_\mu(\pq_1)\varphi^*_\mu(\pq_2)
\times \sum_{\nu}\varphi_\nu^*(\pq_1)\varphi_\nu(\pq_2).  
\end{eqnarray} 
The contribution from both the diagonal ($\mu=\nu$) and 
off-diagonal terms is small: The former is of order $ N^{-1}$, 
while the latter involves $N^2$ terms each of which is of order $N^{-2}$. 
However, since the the phases of wave functions at different point 
are almost uncorrelated so the off-diagonal terms arise with random 
phases. This implies a contribution of the off-diagonal terms which 
is also of order $N^{-1}$. Another way to see this is by invoking 
the completeness argument: Each sum in the last line of 
Eq.~(\ref{nulocmassive}) tends to $\delta( \pq_1 - \pq_2)$ as 
the width of the band is increased (this follows from completeness of the 
basis of eigenstates of the Hamiltonian). At any finite band width 
the function $\sum_{\mu}\varphi_\mu(\pq_1)\varphi^*_\mu(\pq_2)$ 
has the characteristic width $(\Omega/N)^{1/d}$, where $\Omega$ is the volume 
of the system, and can be approximated by the Heaviside function
$ (N/\Omega) \Theta((\Omega/N)^{1/d} - |\pq_1 - \pq_2 |)$. 
Therefore the r.h.s. of Eq.~(\ref{nulocmassive}) vanishes 
if the coordinates $\pq_1$ and $\pq_2$ are separated by a distance 
larger than $(\Omega/N)^{1/d}$. These considerations enable 
us to neglect the massive modes.  

The integration measure in Eq.~(\ref{genfunc}) is invariant under 
the group of transformations $\hat{Q}\to \hat{U}^{-1}\hat{Q}\hat{U}$, 
where $\hat{U}$ is an operator satisfying Eq.~(\ref{group}) 
with indices both in the Hilbert space $U_{\mu \nu}$ and in superspace. 
The action Eq.~(\ref{efftheory}) is 
also invariant under such transformations, provided that $\hat{U}$ 
commutes with $\hat{H}$. We will denote such transformations by $\hat{U}_0$
This symmetry leads to the existence of a degenerate manifold of saddle-point 
solutions (at $s=0$). All matrices of the form 
\begin{equation}
\label{Qmassless}
\hat{Q}= \hat{U}_0^{-1}\hat{Q}_0 \hat{U}_0,  
\end{equation} 
where $[\hat{U}_0,\hat{H}]=0$ satisfy  
Eq.~(\ref{saddlept}). In the basis of the eigenstates of the 
Hamiltonian such matrices are of the form $U_{0,\mu \nu}=\delta_{\mu \nu}
U_{0,\mu}$ with $U_{0,\mu}\in UOSP(2,2/4)$~\cite{Verbaarschot85}. 
We assume the absence of degeneracies due to non-integrability. 
All such matrices generate  
zero-modes. 

It is shown below that the integration over the massive modes strongly 
favors the ground state configurations of $Q$ which correspond to 
identical $Q_\mu$'s. This happens because the ground state in which 
$Q_\mu$'s are different breaks supersymmetry of the action for 
the massive modes. This leads to a rapid decay (as a function of 
inhomogeneity of $Q_\mu$) of the superdeterminant which arises 
from the integration over the massive modes. 
Therefore the integration over the massive modes 
gives a non-vanishing contribution to the effective action which 
depends on $Q_\mu$'s. This contribution can be interpreted as an effective 
interaction between $Q_\mu$'s which favors configurations with 
identical $Q_\mu$'s. Hence, it can be thought of as ``ferromagnetic'' 
interaction of ``spins'' $Q_\mu$ which reside on the nonlinear 
manifold $UOSP(2,2/4)/[UOSP(2/2)\times UOSP(2/2)]$. This 
interaction is long range (all ``spins'' within the band 
interact with approximately equal strength) 
and therefore, in the thermodynamic limit $N\to \infty$, leads to 
a ferromagnetic ground state. The fluctuations of ``spins'' from the 
the ground state configurations are small as $1/N$ and can be neglected. 

To see how the supersymmetry breaking for the massive modes arises let us 
consider one term in the sum (\ref{fluct}) corresponding to 
particular $\mu$ and $\nu$. The matrix $Q_\mu$ has the same 
symmetries as the $Q$-matrix in Efetov's nonlinear 
$\sigma$-model and can be parametrized as
\begin{eqnarray}
  \label{Qefet}
  Q_\mu &=& \left(\begin{array}{cc}
u_\mu & 0 \\
0 & v_\nu 
\end{array}\right)
Q_E(\hat{\theta}_\mu)
\left(\begin{array}{cc}
\bar{u}_\mu & 0 \\
0 & \bar{v}_\mu 
\end{array}\right) , \nonumber \\
Q_E(\hat{\theta}_\mu)&=&\left(\begin{array}{cc}
\cos(\hat{\theta}_\mu) & i\sin(\hat{\theta}_\mu) \\
-i\sin(\hat{\theta}_\mu)) & -\cos(\hat{\theta}_\mu)
\end{array}\right), \qquad \hat{\theta}_\mu ={\rm diag}(\theta, 
\theta,i \theta^+, i\theta^-).
\end{eqnarray}
Here we deviate from Efetov's original parametrization by
introducing the angles $\theta^+$ and $\theta^-$ which can be expressed 
through the angles appearing in Ref.~\cite{Efetov83}, $\theta_1$ and 
$\theta_2$ as $\theta^+=\theta_1+\theta_2$ and $\theta^-=\theta_1-\theta_2$. 
The particular form of the matrices $u$ and $v$ is not important for what 
follows and will be left unspecified.

If the angles $\hat{\theta}_\mu$ and $\hat{\theta}_\nu$ coincide then the 
massive modes line up with $\hat{\theta}_\nu$. In other words we can make a 
global rotation to bring $\hat{\theta}_\nu$ to zero, and in this coordinate 
frame the massive fluctuations correspond to $\delta Q_{\mu \nu}^{RR}$
and $\delta Q_{\mu \nu}^{AA}$. If the angles $\hat{\theta}_\mu$ and 
$\hat{\theta}_\nu$ differ by a small amount, we can go to the 
``center of mass'' coordinate where 
\begin{equation}
  \label{centmass}
  \hat{\theta}_\mu=-\hat{\theta}_\nu.
\end{equation}
In this frame the massive modes will still correspond to $\delta 
Q_{\mu \nu}^{RR}$ and $\delta Q_{\mu \nu}^{AA}$. The contribution of $\delta 
Q_{\mu \nu}^{AA}$ to the effective action is 
\begin{equation}
  \label{fluctmunu}
  {\rm STr}\left(Q_E(\hat{\theta}_\mu) \bar{u}_\mu \delta 
Q_{\mu \nu}^{AA}
u_\nu Q_E(\hat{\theta}_\nu)\bar{u}_\nu \delta 
Q_{\nu \mu}^{AA}u_\mu
+ \delta Q_{\mu \nu}^{AA}\delta Q_{\nu \mu}^{AA}\right) 
\end{equation}

Instead of integration variables $\delta Q_{\mu \nu}^{AA}$ and 
$\delta Q_{\nu \mu}^{AA}$ it is more convenient to use 
$\delta \tilde{Q}_{\mu \nu}^{AA}=\bar{u}_\mu 
\delta Q_{\mu \nu}^{AA}u_\nu$ and $\delta 
\tilde{Q}_{\nu \mu}^{AA}=
\bar{u}_\nu \delta Q_{\nu \mu}^{AA}u_\mu$. Since the superjacobian of such 
transformation is equal to unity,
\begin{equation}
{\rm SDet}\left(\frac{\partial (\delta Q_{\mu \nu}^{AA}, 
\delta Q_{\nu \mu}^{AA})}{\partial (\delta \tilde{Q}_{\mu \nu}^{AA},\delta 
\tilde{Q}_{\nu \mu}^{AA}) }\right)=1,
\end{equation}
the invariant measure is preserved.

With the parametrization involving ordinary variables $a_i$, $b_i$, and
Grassmann variables $\sigma_i$, $\sigma^*_i$,
\begin{equation}
  \label{massiveparam}
  \delta \tilde{Q}_{\mu \nu}^{AA}=\left(\begin{array}{cccc}
a_1 & a_2 & i \sigma_1 & i\sigma_2\\
-a_2^* & a_1^* & -i \sigma_2^* & -i\sigma_1^*\\
 \sigma_3^* & \sigma_4 & ib_1 &  ib_2 \\
 \sigma_4^* & \sigma_3 & ib_2^* &  ib_1^* 
\end{array}\right), \qquad 
\delta \tilde{Q}_{\nu \mu}^{AA}=\left(\begin{array}{cccc}
a_1^* & -a_2 & - \sigma_3 & -\sigma_4\\
a_2^* & a_1 &  \sigma_4^* & \sigma_3^*\\
-i \sigma_1^* & -i\sigma_2 & ib_1^* &  ib_2 \\
-i \sigma_2^* & -i\sigma_1 & ib_2^* &  ib_1 
\end{array}\right),
\end{equation}
which obey the symmetry relations
\begin{equation}
  \label{massivesymmetry}
  \delta \tilde{Q}_{\mu \nu}^{AA}=C^T (\delta 
\tilde{Q}_{\nu \mu}^{AA})^T C, 
\qquad \delta \tilde{Q}_{\mu \nu}^{AA} = k (
\delta \tilde{Q}_{\nu \mu}^{AA})^\dagger,
\end{equation}
integration over massive modes $\delta {Q}_{\mu \nu}^{AA}$ 
and $\delta {Q}_{\nu \mu}^{AA}$ with the action Eq.~(\ref{fluctmunu})
can be performed and yields
\begin{equation}
  \label{superdetmass}
  I_{\mu \nu}= \frac{[2+\cos\theta (\cosh\theta^+ +\cosh\theta^-)]^4}{
(2+\cosh^2\theta^+ +\cosh^2\theta^-)(2+2\cosh\theta^+\cosh\theta^-)
(2+2\cos^2\theta)^2}.
\end{equation}
For small $\hat{\theta}_\mu$, writing $\cos\theta =1-\alpha$, 
$\cosh\theta^+=1+\beta^+$, and $\cosh\theta^-=1+\beta^-$, $I_{\mu \nu}$ 
can be expanded to second order,
\begin{equation}
I_{\mu \nu}\approx 1-{1\over 8}(2\alpha +\beta^+ + \beta^-)^2 \approx 
\exp\left[-{1\over 8}(2\alpha +\beta^+ + \beta^-)^2\right].
\end{equation} 

If all $\hat{\theta}_\mu$ are small, then we obtain a model equivalent to 
spins with infinite range interactions. In the thermodynamic limit of 
such a model the mean field approximation becomes exact. The fluctuations of 
$\alpha$, $\beta^+$ and $\beta^-$ become small as $1/N$ and can be neglected.
This forces us to consider the matrices $\hat{Q}_0$ which are of the form 
of Eq.~(\ref{spsolutdiag}). Then the relevant (massless) fluctuations of the 
$Q$-matrix are those that anticommute with $\Lambda$ in superspace.

\section{Perturbation theory}
\label{perturbation}

In this appendix we will employ the $\sigma$-model to study to the two-point 
correlator of DoS fluctuations. In particular, we will examine the 
perturbative corrections to RMT which appear at larger energy scales. 
To obtain the high frequency asymptotics of $R_2(s)$ the functional integral 
in Eq.~(\ref{r2gen}) can be evaluated using the stationary point 
approximation. However, to obtain the contribution which is non-perturbative 
in $1/s$ it is necessary to introduce an additional term $u^2(\Lambda Q)^2$ 
into the action (\ref{ZR2}) which serves as a regulator controlling the 
stationary point approximation~\cite{Andreev95,Jona-lasinio96}. Ultimately, 
the regularization parameter $u$ can be set to zero.

We therefore express the two-point correlator as 
\begin{equation} 
R_2(s)=\lim_{u\to 0}\frac{1}{64}\Re\int {\cal D}Q 
\left( \int d \px_\parallel {\rm STr} 
[\Lambda k Q(\px_\parallel)]\right)^2
\exp\left[-S_{\em eff}(s)\right],
\label{r2gen1} 
\end{equation} 
where  
\begin{eqnarray}
  \label{ZR21}
  S_{\em eff}(s)&=& 
 {\pi\over 2}\int d\px_\parallel
 {\rm STr} \left[i\frac{s^+}{2} \Lambda Q -u^2(\Lambda Q)^2
+QT^{-1}\hat{\cal L} T\right].
\end{eqnarray}
The derivation of the results already presented in Eqs.~(\ref{perturbative}) 
and (\ref{ZR2pert}) closely parallels that of Ref.~\cite{Andreev95}. For a 
more detailed account of the method see Ref.~\cite{Jona-lasinio96}.
At high frequency the integrand in Eq.~(\ref{ZR21}) becomes highly 
oscillatory, and we can use the stationary phase method to evaluate the 
integral. We will show that there are two stationary points: $Q=\Lambda$ and 
$Q=-\Lambda k$. The term $u^2(\Lambda Q)^2$ in the action is introduced in 
order to stabilize the second one. We can expand the integrand in small 
fluctuations of the $Q$-matrix around $\Lambda$ and $-\Lambda k$ to obtain the 
leading high-frequency asymptotics of $R_2(s)$. 

We first consider the expansion around $Q=\Lambda$. This corresponds to the 
ordinary perturbation expansion previously employed in the study of disordered 
conductors~\cite{Efetov83,Altshuler86}. We begin with the parametrisation
\begin{equation}
   \label{P}
  T = 
\openone +i P, \qquad P=\left(\begin{array}{cc}
0 & B \\
\bar{B} & 0
\end{array}\right), 
\end{equation}
where, from Eq.~(\ref{Tcharge}), it follows that $P$ satisfies the condition  
\begin{equation}
  \label{Pcharge}
  P(\pq,\pp)^*=-CP(\pq,-\pp)C^T.
\end{equation}

Next we substitute Eq.~(\ref{P}) into Eq.~(\ref{r2gen1}) and expand the 
integrals in the pre-exponential factor and the free energy (\ref{ZR21})
to second order in $P$. Due to the presence of the infinitesimal imaginary
part in $s^+$, the stationary point $Q=\Lambda$ is stable and we can safely 
set $u=0$ in the free energy (\ref{ZR21}). 
To second  order in $B$ and $\bar{B}$ we have 
\begin{mathletters}
\begin{eqnarray}
\label{Qexpan}
  {\rm STr}(\Lambda kQ)&\approx& 
8-2{\rm STr}(kB\bar{B}+k\bar{B}B), \qquad 
{\rm STr}(\Lambda kQ)^2\approx 
-8{\rm STr}(k\bar{B}kB+\bar{B}B)
\\
 {\rm STr}(\Lambda Q)& \approx & -4{\rm STr}(\bar{B}B).
\end{eqnarray}
\end{mathletters}
Using these relations we obtain the following expression for the 
perturbative part of $R_2(s)$
\begin{eqnarray}
  \label{ptaction}
R_{P}(s)&=& \Re\int {\cal D}[B,\bar{B}] 
\left(\int d \px_\parallel 
[1-\frac{1}{2}{\rm STr} (kB\bar{B}+k\bar{B}B)]\right)^2 
\exp\left[-S_{{\em eff}}(s) \right],
\end{eqnarray}
where 
\begin{equation}
  \label{freenpert}
 S_{{\em eff}}(s)=i\pi \int (d\px_\parallel)\ {\rm STr}  
\left[- s\bar{B}B -i
\bar{B}\hat{\cal L} B\right] + O(B^4).
\end{equation}

In order to perform the integration over $B$ and $\bar{B}$
it is convenient to represent these matrices as 
\begin{equation}
  \label{Bcooperdif}
  B=\sum_{i=0}^3 B_i \tau_i, \qquad \bar{B}=\sum_{i=0}^3 \bar{B}_i \tau_i,
\end{equation}
where the Pauli matrices $\tau_i$ are defined in Eq.~(\ref{Paulimat}).
As follows from Eq.~(\ref{groupwig}) the matrices $B$ and $\bar{B}$ 
in Eq.~(\ref{P}) obey the relation $\bar{B}=kB^\dagger$, which implies 
\begin{equation}
  \label{Bpm}
  \bar{B}_i=kB_i^\dagger,  \quad i=0,\ldots ,3.
\end{equation}
In this notation, Eq.~(\ref{freenpert}) becomes 
\begin{equation}
  \label{ptactionpm}
    S_{{\em eff},P}(s)= - i \pi \int (d\px_\parallel)\ {\rm STr}  
\left[\sum _{i=0}^3\left( \bar{B}_i(s +i \hat{\cal L}) B_i \right)\right].
\end{equation}

Each matrix $B_i$ can be parametrized as 
\begin{equation}
  \label{Biparam}
  B_i=\left( \begin{array}{cc}
a_i & i\sigma_i \\
\eta_i^* & i b_i
\end{array}\right).
\end{equation}
The parametrization for $\bar{B}_i$ can be obtained from Eq.~(\ref{Bpm}).
To evaluate the integral (\ref{ptaction}) over the variables 
(\ref{Biparam}) one can use Wick's theorem. It is necessary to take into 
account Eq.~(\ref{Pcharge}) which reduces the number of independent 
integration variables by a factor of two. As a result we obtain the 
second part in Eq.~(\ref{perturbative}).

The stationary point $Q=\Lambda$ of the functional integral 
(\ref{r2gen1}) is not the only one. To find the other 
stationary points consider Eq.~(\ref{r2gen1}). 
It is possible to parameterize fluctuations around a general 
stationary point $Q_0$ as $Q=Q_0(1+iP_0)(1-iP_0)^{-1}$, 
where $P_0$ anticommutes with $Q_0$ and  no longer 
obeys equations (\ref{P}) and (\ref{Pcharge}). 
Expanding the effective action in Eq.~(\ref{ZR21}) in powers of 
$P_0$ we would obtain the stationarity condition 
$\partial S_{\em eff}(s)/\partial P_0 =0$. 

This route however is inconvenient since the parametrization of $P_0$ 
will depend explicitly on $Q_0$. Instead it is convenient to perform a 
{\em global} coordinate transformation on ${\bf H}=U(2,2/4)/[U(2/2)\times
U(2/2)]$, $Q\to U_0^{-1}Q U_0$, where $U_0 \in {\bf H}$, which maps 
$Q_0$ to $\Lambda$.

Since all points on a symmetric 
space are equivalent by definition, this coordinate transformation preserves 
the invariant measure and leaves the functional integral in 
Eq.~(\ref{r2gen1}) invariant. The integrand, however, will change because 
it contains matrices $\Lambda$ and $-k\Lambda$ that break the symmetry 
in the coset space. Such a coordinate transformation is equivalent 
to changing only the source matrices $\Lambda 
\to Q_\Lambda = U_0 \Lambda U_0^{-1}$ and $-k\Lambda \to - Q_{k \Lambda}
= - U_0 k \Lambda U_0^{-1}$ in Eqs.~(\ref{r2gen1}), (\ref{ZR21}) and keeping 
the old parametrization of Eq.~(\ref{P}). The stationary points will 
correspond to those $U_0$
for which the linear in $P$ terms in the expansion of the 
effective action vanish. 

Note that because the transformation matrix $U_0$ is 
independent of momenta $\pp$, from Eq.~(\ref{Uchargewign}) 
it follows that $U_0^*=C U_0 C^T $. Therefore $U_0$ belongs to the 
coset space of the usual orthogonal ensemble~\cite{Verbaarschot85}, 
$UOSP(2,2/4)/[UOSP(2/2)\times UOSP(2/2)]$. For this case there 
is only one other stationary point~\cite{Andreev95,Jona-lasinio96}
corresponding to 
$Q_\Lambda =  U_0 \Lambda U_0^{-1}=-k\Lambda$ and 
$-Q_{k\Lambda} = - U_0 k \Lambda U_0^{-1}= \Lambda$.
For this point Eq.~(\ref{r2gen1}) can be rewritten as 
\begin{mathletters}
\begin{eqnarray} 
R_2(s)&=&-\lim_{u\to 0}\frac{1}{64}\Re\int {\cal D}Q 
\left( \int d \px_\parallel {\rm STr} 
[\Lambda  Q(\px_\parallel)]\right)^2
\exp\left[-\tilde{S}_{{\em eff}}(s)\right],
\label{r2genk} 
\\ 
  \label{ZR2k}
  \tilde{S}_{{\em eff}}(s,u)&=& 
 {\pi\over 4}\int (d\px_\parallel) 
 {\rm STr} \left[-is^+ k\Lambda Q -u^2(k \Lambda Q)^2
+QT^{-1}\hat{\cal L} T\right].
\end{eqnarray}
\label{r2genkab}
\end{mathletters}
We now expand Eq.~(\ref{r2genk}) in powers of $P$ using Eq.~(\ref{P}). 
This expansion is equivalent to expanding the $Q$-matrix around 
$-k\Lambda$ in Eq.~(\ref{r2gen1}). Expanding the free energy (\ref{ZR2k}) to 
second order in $P$ we use Eq.~(\ref{Qexpan}). 
Note that with the parametrization of Eq.~(\ref{Biparam})
\begin{mathletters}
\begin{eqnarray}
\label{sourceactiona}
  {\rm STr} \left( k\bar{B} B + k B \bar{B} \right)&=& 
-4\sum_{i=0}^3 (|a_i|^2 - |b_i|^2), \\ 
\qquad {\rm STr} \left( k\bar{B}k B +  \bar{B} B \right)
&=&-4 \sum_{i=0}^3 (|a_i|^2 + |b_i|^2).
\label{sourceactionb}
\end{eqnarray}
\label{sourceaction}
\end{mathletters}
Therefore the Grassmann variables in the parametrization (\ref{Biparam}) 
do not couple to $s$ and $u^2$. As follows from Eq.~(\ref{sourceactiona}), 
the ordinary variables $a_i$ and $b_i$ couple to $s$ 
with opposite sign. Due to the presence of the infinitesimal imaginary 
part in $s$ the integral over $a_i^0$ (the zero mode variable) would diverge
at $u=0$. Eq.~(\ref{sourceactionb}) shows that the term 
${\rm STr}(k\Lambda Q)^2$ makes the integration over $a_i^0$ convergent.
We therefore have to keep $u$ finite during the evaluation of the 
functional integral and take the limit $u \to 0$ only in the final 
expressions. The quadratic approximation to the free energy (\ref{ZR2k}) 
becomes
\begin{equation}
  \label{ZR2kpert}
  \tilde{S}_{{\em eff}}(s)= -2 \pi i s +
  2\pi \int (d\px_\parallel) 
 {\rm STr} \left[ 
\sum_{i=0}^3\left(a_i^* (-is^+  -u^2 + \hat{\cal L})a_i 
+b_i^* (is^+  -u^2 +\hat{\cal L})b_i 
+\sigma_i^*  \hat{\cal L}\sigma_i 
+\eta_i^*  \hat{\cal L}\eta_i
\right)\right].
\end{equation}
The zero  mode Grassmann variables $\eta_i^0$ and $\sigma_i^0$ do 
not appear in the quadratic expansion of the effective action 
(\ref{ZR2kpert}). 
For the integral (\ref{r2genk}) not to vanish they have to come from 
the pre-exponential factor. While evaluating the integral we have to take
into account the symmetry (\ref{Pcharge}) which reduces the number 
of independent integration variables by factor of two. Therefore there 
are eight independent Grassmann variables in the zero mode. Thus, in 
order to obtain a non-zero result we should expand the pre-exponential 
factor to eighth order in $P$. Then in the eighth order expansion 
of the prefactor we should keep only the zero mode terms. This renders 
the integration over the zero mode variables non-vanishing, whereas 
the integration over the ordinary zero mode variables yields a factor 
$(s^2+ u^4)^{-2}$. The integral over the non-zero modes yields the 
superdeterminant of the operator (\ref{ZR2kpert}). After we perform 
the integration we take the $u\to 0$ limit to obtain Eq.~(\ref{ZR2pert}).

\end{document}